\documentclass[usenatbib]{mn2e}
\pdfoutput=1
\usepackage[pdftex]{graphicx}
\def\lsim{~\rlap{$<$}{\lower 1.0ex\hbox{$\sim$}}}
\def\gsim{~\rlap{$>$}{\lower 1.0ex\hbox{$\sim$}}}

\begin{document}

\title[The formation of disks and SMBHs]{Gas infall into atomic cooling haloes: on the formation of protogalactic disks and supermassive black holes at $ z > 10$}

\author[Prieto, Jimenez \& Haiman]{Joaquin Prieto$^1$\thanks{email:joaquin.prieto.brito@gmail.com}, Raul Jimenez$^{2,1,3}$, Zolt\'{a}n Haiman$^4$\\
$^{1}$ICC, Universitat de Barcelona (IEEC-UB), Marti i Franques 1, E08028, Barcelona, Spain\\
$^2$ ICREA\\
$^3$ Theory Group, Physics Department, CERN, CH-1211, Geneva 23, Switzerland\\
$^2$ Department of Astronomy, Columbia University, 550 West 120th Street, MC 5246, New York, NY 10027, USA}

\maketitle

\begin{abstract}
  We have performed hydrodynamical simulations from cosmological
  initial conditions using the AMR code RAMSES to study atomic cooling
  haloes (ACHs) at $z=10$ with masses in the range $5\times10^7{\rm
    M_{\odot}}\lsim\, M\lsim 2\times 10^9{\rm M_{\odot}}$.  We assume
  the gas has primordial composition and ${\rm H_2}$-cooling and prior
  star-formation in the haloes have been suppressed.  We present a
  comprehensive analysis of the gas and DM properties of 19 haloes at
  a spatial resolution of $\sim10$ (proper) pc, selected from
  simulations with a total volume of $\sim 2000$ (comoving) Mpc$^3$.
  This is the largest statistical hydro-simulation study of ACHs at
  $z>10$ to date.  We examine the morphology, angular momentum,
  thermodynamical state, and turbulent properties of these haloes, in
  order to assess the prevalence of disks and massive overdensities
  that may lead to the formation of supermassive black holes (SMBHs).
  We find no correlation between either the magnitude or the direction
  of the angular momentum of the gas and its parent DM halo.  Only $3$
  of the haloes form rotationally supported cores.  Two of the most
  massive haloes, however, form massive, compact over-dense blobs,
  which migrate to the outer region of the halo.  
  These blobs have an accretion rate $\sim{\rm 0.5M_{\odot} yr^{-1}}$
  (at a distance of 100 pc from their center), and are possible sites
  of SMBH formation.  Our results suggest that the degree of
  rotational support and the fate of the gas in a halo is determined
  by its large-scale environment and merger history.  In particular,
  the two haloes that form over-dense blobs are located at knots of the
  cosmic web, cooled their gas early on ($z>17$), and experienced many
  mergers.  The gas in these haloes is thus lumpy and highly turbulent,
  with Mach numbers ${\mathcal M}\ga 5$.  In contrast, the haloes
  forming rotationally supported cores are relatively more isolated,
  located midway along filaments of the cosmic web, cooled their gas
  more recently, and underwent fewer mergers.  As a result, the gas in
  these haloes is less lumpy and less turbulent (Mach numbers
  ${\mathcal M}\la 4$), and could retain most of its angular
  momentum.  The remaining 14 haloes have a diverse range of
  intermediate properties. If verified in a larger
  sample of haloes and with additional physics to account for metals
  and star-formation, our results will have implications for
  observations of the highest-redshift galaxies and quasars with {\rm JWST}.
\end{abstract}


\section{Introduction}

According to the paradigm of hierarchical structure formation, the
dissipationless dark matter (DM) particles build up structures in the
Universe from small building blocks to create increasing more massive
DM haloes. Due to its ability to cool radiatively, the baryonic matter
can loose pressure support and condense inside the DM haloes, forming
the first luminous objects: stars, galaxies, and black holes (BHs).

The formation of the first stars has been extensively studied in the
last decade by a number of authors
(e.g. \citet{Abeletal2000,Yoshidaetal2006,Turketal2012}). These
studies have shown that the first stars (pop III stars) were formed
due to molecular hydrogen cooling inside DM haloes of $\sim10^6{\rm
  M_{\odot}}$ at redshift as high as $z\sim30$. Numerical simulations
of the first stars were extended to the DM haloes in the mass range
$\sim(10^7-10^8){\rm M_\odot}$ DM by e.g. \citet{WiseAbel2007},
\citet{Greifetal2008}, and \citet{jpp3} in order to study the physical
processes involved in the formation of the first
galaxies. Unfortunately, individual pop III stars (or even clusters of
hundreds of these stars) are too faint to be detected even with
current (and future) observations.
 
The possibility that the {\it James Web Space Telescope} ({\it JWST})
can observe the first galaxies, located in haloes $\gsim 10$ times more
massive than those hosting the first PopIII stars, is an exciting
motivation to study the so-called atomic cooling haloes (i. e. DM
haloes with a virial temperature $T_{\rm vir}\ga10^4K$; hereafter ACHs
-- see e.g. \citet{BrommYoshida2011,Ferrara} for a recent reviews).  These types
of haloes have been studied recently by \citet{RomanoDiaz2011} and
\citet{Pawliketal2011} in the context of galactic disk formation with
smoothed particle hydrodynamic (SPH)
simulations. \citet{RomanoDiaz2011} presented a low-resolution
statistical study of haloes with masses above $\sim10^8{\rm
M_{\odot}}$ inside a highly biased over-dense region at $z=10$, showing
that every halo with mass above $\sim10^9{\rm M_{\odot}}$ develops a
gas disk structure. \citet{Pawliketal2011} have studied the formation
of a galactic disk inside a single $10^9{\rm M_{\odot}}$ halo using a
high-resolution SPH simulation at the same redshift.
\citet{Wiseetal2012} similarly studied the formation of two
proto-galaxies in high-resolution AMR simulations that also included
star-formation and chemical and radiative feedback.
Protogalactic disk formation inside ACHs has been studied using
analytical (semi-phenomenological) approaches as well. Under the
simple assumption \citep{Moetal1998} that the disk specific angular
momentum is a fixed fraction of the DM halo specific angular momentum,
\citet{OhHaiman2002} found that most haloes are able to form a stable
disks (the exceptions are haloes with very low spin parameter).
\citet{StringerBenson2007} extended the galactic disk study, taking
into account the hierarchical nature of the structure formation. Under
the assumption of conservation of specific angular momentum during the
collapse process they studied the effect of mergers on disk formation,
concluding that mergers have a significant impact on the prevalence of
disks.

Pristine ACHs have been proposed as possible sites for the formation
of supermassive black holes (SMBHs).  If the gas were able to remain at a
temperature of $\sim 10^4$K in these haloes, it has been proposed that
fragmentation may be avoided, allowing the formation of a $\sim
10^{5-6}~{\rm M_\odot}$ supermassive black hole (SMBH) at the nucleus
\citep{OhHaiman2002,Koushiappas+2004,VolonteriRees2005,LodatoNatarajan2006,SpaansSilk2006,Begelman+2006}.
These SMBHs could then grow by another factor of $10^{3-4}$ by
redshift $z\sim 6$ and explain the puzzling presence of $M\gsim 10^9
{M_\odot}$ BHs by those redshifts (see, e.g., the recent review in
\citealt{Haiman2012}).  A few groups have now performed simulations of
gas collapse in such haloes
\citep{Wiseetal2008,ReganHaehnelt2009a,ReganHaehnelt2009b,Latifetal2011},
with the simplifying assumption that ${\rm H_2}$ cooling can be
avoided, which may be feasible in case the haloes are illuminated by a
strong Lyman Werner background
\citep{Omukai2001,BrommLoeb2003,Shangetal2010}.  These works have
indeed found no evidence of fragmentation.

In this paper we continue the study of ACHs using AMR simulations. Our
work is motivated by the possible formation of galactic disks and
SMBHs inside these haloes, as well as by the fact that these haloes
may host the first galaxies that will be directly detectable by {\it
JWST}.  Our study is the largest in the literature based on
hydrodynamical simulations, with a total of 19 haloes analysed at a
spatial resolution of $\Delta x\sim10$ pc.  We present, for the first
time, a comprehensive study of gas and DM physical properties,
including their spatial distribution, angular momenta, thermodynamical
state, and turbulent properties.  We go beyond commonly used
diagnostic tools, such as spherically averaged profiles, and compute
density and velocity power spectra, and density, velocity and
temperature probability distribution functions.  Using our sample of
haloes, we attempt to identify the physical mechanisms that determine
the angular momentum of the gas in the inner halo, and therefore
whether a rotationally supported inner disk or a low-spin, compact
dense cusp (and ultimately a SMBH) forms.

Although we find essentially no correlation between the magnitude or
direction of the angular momenta of the gas and the DM haloes, there
are clear differences between the haloes hosting the two types of
inner gas: haloes with low-spin central cusps form in highly biased
regions at super-knots of the cosmic web, and have cooled their gas
early on and undergone many mergers. This results in a lumpy and
highly turbulent medium that could produce efficient transfer of
angular momentum. On the other hand, the rotationally supported core 
is found in haloes that are located midway along cosmic
filaments, cooled relatively more recently, and have experienced fewer
mergers.  The gas in these haloes is less lumpy and less turbulent and
could retain most of its angular momentum.  By a detailed study of the
turbulent properties of the intra-halo medium (hereafter IHM) we have
found that turbulence plays a major role at dissipating (or
preserving) angular momentum via fragmentation of the IHM. Turbulence
in turn is fed by the merging process. We will show that compact
over-dense blobs (COBs) form in highly biased regions at the centers of
super-knots, while rotationally supported cores (RSCs) form in the
middle of filaments in nearly un-biased regions.  Our numerical
experiments suggest that neither rotationally supported disks nor
SMBHs are a common occurrence in ACHs, and should be followed up by
studying a larger sample of haloes, with more realistic gas physics and
more detailed analyses of turbulent angular momentum transport.

The paper is organized as follows: In section {\S}\ref{Methodology} we
describe the details of our numerical simulations and our halo
sample. In {\S}\ref{Results}, we explain our analysis and show the
physical quantities extracted from the simulations. In section
{\S}\ref{Discussion}, we discuss our results and finally in
{\S}\ref{Conclusions} we present our conclusions.

\section{Numerical Simulation Details}
\label{Methodology}

We used the AMR code RAMSES \citep{Teyssier2002} with a modified
non-equilibrium cooling module \citep{jpp} in order to study the main
physical properties of both DM and gas components of 19 ACHs in the
mass range $\sim 5\times 10^{7}$M$_\odot$ to $\sim2 \times
10^{9}$M$_{\odot}$. We assumed the gas has primordial composition, and
switched off all molecular cooling processes, allowing the gas to cool
down only due to atomic hydrogen and helium lines.

We worked in a concordance $\Lambda$CDM cosmological model: $h=0.719$,
$\Omega_{\Lambda}=0.742$, $\Omega_{m}=0.258$, $\Omega_{b}=0.0441$,
$\sigma_8=0.796$ and $n_s=0.963$ \citep{Komatsu2011}.  The dynamical
initial conditions were taken from {\em mpgrafic}
\citep{Prunetetal2008} and the initial abundances of H, H$^+$, He,
He$^+$ and He$^{++}$, D, D$^+$ and $e^-$ from \citet{GP98} at $z=100$
(the initial redshift of the simulations).

First, we performed a number of DM-only simulations with 256$^3$
particles inside (3 Mpc)$^3$, (5 Mpc)$^3$, (5.5 Mpc)$^3$ and (8
Mpc)$^3$ boxes in order to look for ACHs, i.e., haloes with virial
temperatures T$_{\mathrm{\rm vir}}\ga 10^4$K, which corresponds to a
$M_{\rm vir}\ga10^8$M$_{\odot}$ at $z=10$.  The relation $T_{\rm
  vir}(M_{\rm vir})$ was taken from \citet{BryanNorman1998} and
$M_{\rm vir}$ refers to the mass for a DM overdensity with a density
contrast $\delta=200$.  In this study, we focus on $z=10$, which
corresponds to the epoch when ACHs are ``2-$\sigma$'' objects and
begin to have a significant abundance (a few per Mpc$^3$).  Using the
HOP algorithm \citep{EisensteinHut1998} we identified DM haloes in the
DM simulations outputs at this redshift.  Our goal was to study the
simplest, relatively isolated first protogalaxy, and we therefore
avoided haloes that are undergoing an obvious merger at $z=10$.  We
have found that mergers are very common at high redshift, and we have
rejected many simulations (containing $\sim 100$ pre-selected
haloes) in which no such isolated DM halo was found, before arriving
at a sample of 19 suitable, non-merging DM haloes.  We then performed a
new set of gas+DM simulations for these 19 haloes, in 19 different
runs.  Each of these re-simulation runs were performed with $512^3$ DM
particles inside the corresponding original simulation volume.
from $z=100$ to $z=10$ in order to follow the main process of gas
collapse into the ACHs at higher resolution.  In Table~\ref{table1},
we summarize the properties of our 19 haloes, and the simulations in
which they were identified.  The virial quantities are computed from
the halo mass taking the expressions of \citet{BryanNorman1998}.  Each
of our simulations used $\sim60,000$ CPU hours (for a total of
$\sim10^6$ total CPU hours for the whole simulation suite) on a cluster with $500$ cores and $2$ TB of distributed memory.

\begin{table*}
\begin{center}
  \caption{Properties of the 19 atomic cooling haloes at $z=10$
    analyzed in this paper, and the simulations in which they were
    identified.  The first 4 columns indicate the id number, mass,
    virial temperature, and virial radius of each halo, ranked by its
    total mass. The 5$^{\rm th}$ column shows the type of gaseous
    object forming in the halo (rotationally supported core or compact
    over-dense blob).  The 6-8$^{\rm th}$ columns show, respectively,
    the comoving box size, DM particle mass, and spatial resolution
    for the gas (in proper units, at the highest level of refinement
    at $z=10$).}
 \begin{tabular}{cccccccc}
\hline\hline
           &                  & & & &          &                   &                       \\
Halo       & Halo mass        & T$_{\rm vir}$ & R$_{\rm vir}$& Object & Box  & m$_{\rm DM}$ & Resolution  \\
(number)   & (M$_{\odot}$)    & (K) & (kpc) & & (Mpc)      & (M$_{\odot}$)       & (pc)                   \\
           &                  & & & &            &                   &                       \\
\hline                                   
           &                  & & & &           &                   &                       \\  
1          & 5.35$\times10^7$ & 2.34$\times10^4$ & 0.72 & RSC & 3.0       & 7.38$\times10^3$  &   8.3                 \\  
2          & 1.02$\times10^8$ & 3.60$\times10^4$ & 0.89 & --   & 3.0       & 7.38$\times10^3$  &   8.3                 \\  
3          & 1.47$\times10^8$ & 4.59$\times10^4$ & 1.00 & --   & 3.0       & 7.38$\times10^3$  &   8.3                 \\ 
4          & 1.56$\times10^8$ & 4.78$\times10^4$ & 1.02 & --   & 3.0       & 7.38$\times10^3$  &   8.3                 \\  
5          & 3.21$\times10^8$ & 7.73$\times10^4$ & 1.31 & --   & 3.0       & 7.38$\times10^3$  &   8.3                 \\ 
6          & 3.39$\times10^8$ & 8.02$\times10^4$ & 1.33 & --   & 3.0       & 7.38$\times10^3$  &   8.3                 \\ 
7          & 3.99$\times10^8$ & 8.94$\times10^4$ & 1.40 & RSC & 5.5       & 4.55$\times10^4$  &   15.3                \\ 
8          & 4.59$\times10^8$ & 9.81$\times10^4$ & 1.47 & --   & 5.0       & 3.42$\times10^4$  &   13.9                \\ 
9          & 4.78$\times10^8$ & 1.00$\times10^5$ & 1.49 & --   & 3.0       & 7.38$\times10^3$  &   8.3                 \\ 
10         & 5.00$\times10^8$ & 1.03$\times10^5$ & 1.51 & --   & 3.0       & 7.38$\times10^3$  &   8.3                 \\ 
11         & 5.40$\times10^8$ & 1.09$\times10^5$ & 1.55 & --   & 5.0       & 3.42$\times10^4$  &   13.9                \\ 
12         & 5.72$\times10^8$ & 1.13$\times10^5$ & 1.58 & RSC & 5.5       & 4.55$\times10^4$  &   15.3                \\ 
13         & 7.81$\times10^8$ & 1.39$\times10^5$ & 1.76 & --   & 3.0       & 7.38$\times10^3$  &   8.3                 \\ 
14         & 8.35$\times10^8$ & 1.46$\times10^5$ & 1.80 & --   & 5.5       & 4.55$\times10^4$  &   15.3                 \\ 
15         & 8.60$\times10^8$ & 1.49$\times10^5$ & 1.81 & --   & 3.0       & 7.38$\times10^3$  &   8.3                 \\ 
16         & 9.22$\times10^8$ & 1.56$\times10^5$ & 1.86 & COB & 3.0       & 7.38$\times10^3$  &   8.3                 \\ 
17         & 1.48$\times10^9$ & 2.14$\times10^5$ & 2.18 & --   & 5.5       & 4.55$\times10^4$  &   15.3                \\ 
18         & 1.53$\times10^9$ & 2.19$\times10^5$ & 2.20 & COB & 5.0       & 3.42$\times10^4$  &   13.9                \\
19         & 1.87$\times10^9$ & 2.50$\times10^5$ & 2.35 & --   & 8.0       & 1.39$\times10^5$  &   22.1                \\  
           &                  & & & &           &                   &                       \\
\hline
\end{tabular}
\label{table1}
\end{center}
\end{table*}

Throughout this study, we adopted the following refinement strategy:
if there were more than $4$ DM particles inside a cell, then the cell
is refined; if the gas density inside a cell is more than $4$ times
the average density inside the whole box, then the cell is refined,
and if the linear cell size is larger than $0.25$ its corresponding
Jeans length, then the cell is refined \citep{Trueloveetal1997}.  The
maximum level of refinement was $l_{max}=15$, corresponding to a
proper resolution at $z=10$ of $8$, $14$, $15$, and $22$pc in the
(3 Mpc)$^3$, (5 Mpc)$^3$, (5.5 Mpc)$^3$ and the (8 Mpc)$^3$ box,
respectively.  Our resolution corresponds to $\sim 1\%$ of the virial
radius and it is not adequate to resolve structures in the inner core
of the halo. We therefore set a gas temperature floor, $T_{\rm
floor}$, in order to avoid numerical fragmentation. This temperature
floor was based on the Truelove criterion that the Jeans length must
be resolved by 4 cells,
\begin{equation}
\lambda_\mathrm{J}=4\Delta x_l,
\end{equation}
where $\lambda_\mathrm{J}$ is the Jeans length, $\Delta x_l$ is the
proper cell size at a given level of refinement $l$.

In their recent study of primordial star formation study with magnetic
field and non-equilibrium ${\rm H_2}$ chemistry, \citet{Turketal2012}
show that a much higher number (64) cells per Jeans length is required
to achieve convergence in the ultra-dense center of the halo.  Since
here we do not include ${\rm H_2}$ chemistry or magnetic fields, and
we are not interested in studying the gas collapse process at sub pc
scales, the standard Truelove criterion should be adequate for our
purposes.  We then obtain the gas temperature floor at a given level
of refinement as
\begin{equation}
T_{\rm floor}=\frac{ 64\pi }{ 15 }\frac{G\mu m_p \rho }{ k_B}\frac{L^2}{2^{2l}},
\end{equation}
where $G$ is the
Newton's gravitational constant, $\mu$ is the mean molecular weight,
$m_p$ is the proton mass, $\rho$ is the gas mass density, $k_B$ is the
Boltzmann constant and $L$ is the proper simulation box size. Assuming
$\mu=1.22$ (appropriate for the nearly neutral $T\lsim 10^4$K
primordial gas in our case) and $l=l_{max}$, $T_{\rm floor}$ at $z=10$
can be written as
\begin{equation}
  T_{\rm
  floor}\approx60\mathrm{K}\left(\frac{n}{1~\mathrm{cm}^{-3}}\right)\left(\frac{L}{5.5~{\rm
  Mpc}}\right)^2
\label{T_floor}
\end{equation}
This temperature floor is much lower than the atomic cooling
temperature threshold $\sim10^4$K, so it does not affect the thermal
evolution of the gas unless its number density exceeds $\ga 100$
cm$^{-3}$.  In most of our haloes, the gas does not reach such high
densities and the temperature floor is not activated. However, there
are two exceptions, where the number density in the core reaches up to
$\sim10^3$cm$^{-3}$.  In these cases the gas temperature in the central
over-dense regions does not have a physical meaning (nevertheless, as
we will see below, our results are sufficient to argue that conditions
in these cores may be most favourable for forming SMBHs).

\begin{figure}
\centering
\includegraphics[height=7.9cm,width=7.8cm]{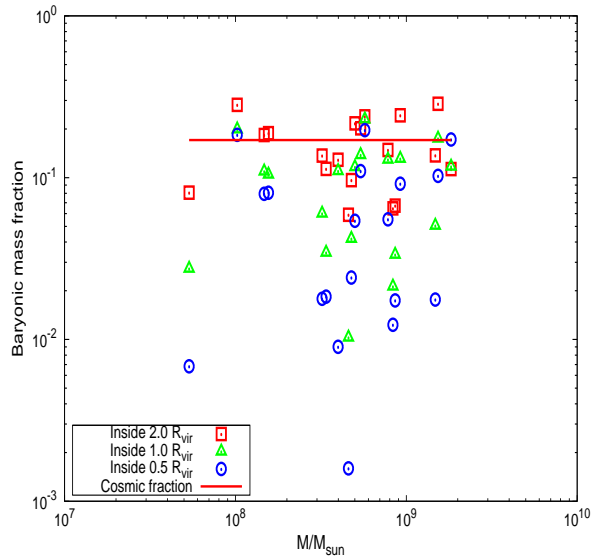}
\caption{Baryon mass fraction inside different radii as a function of
  the halo mass. The horizontal line shows the average cosmic value of
  $\Omega_b/\Omega_m$=0.171 (using the parameters in the text). Circles,
  triangles, and squares show the mass fractions inside 0.5, 1, and
  2$R_{\rm vir}$, respectively.}
\label{b_frac}
\end{figure}

\section{Results}
\label{Results}

In this section, we present an exhaustive list of physical quantities
extracted from our 19 simulated haloes.  These include traditional
global measures (such as gas mass fractions, as well as the
geometrical shapes and spin parameters for both the DM and the gas),
radial profiles (density, temperature, and angular momentum),
three--dimensional morphologies, as well as phase diagrams for the
gas.  In addition, we present several less commonly used diagnostics,
such as probability distribution functions (gas density, temperature
and velocity), and a Fourier analysis (including power spectra of gas
density and velocity), which will help us to understand the physical
conditions in ACHs at their collapse at $z=10$.

\subsection{Gas fractions and phase diagrams}

We computed the baryon mass fraction $f_{b}\equiv M_{\rm gas}/(M_{\rm
DM}+M_{\rm gas})$ inside each of the 19 haloes at three different
radii: 0.5, 1, and 2$R_{\rm vir}$. Here the virial radius $R_{\rm
vir}$ is measured from the center of mass of the DM halo.  Figure
\ref{b_frac} shows these mass fractions as a function of the halo
mass. The average baryon mass fractions are
$(f_{b,0.5},f_{b,1.0},f_{b,2.0})=(0.066, 0.097, 0.16)$, which
correspond to 39\%, 57\% and 92\% of the cosmic average value
($f_b=0.171$).  This is lower than the observed average value found by
\citet{RomanoDiaz2011} at the same redshift and for the same range of
masses using SPH simulations. Our $f_{b}$ value is also lower than
values typically found in galaxy cluster simulations at low redshifts
(e.g., \citet{Gottlober2007}, \citet{Ettori2006} and
\citet{Kravtsov2005}).

The low baryonic mass fractions we find might be partly due to the
fact that the gas temperature is not too far below the virial
temperature, so that gas pressure can affect the dynamics and delay
the collapse of the gas.  Interestingly, however, the mass fractions
lack any obvious trend with halo mass (or with $T_{\rm vir}$), as one
would expect. The lack of this trend suggests that turbulence,
generated at the time of virialization \citep{WiseAbel2007}, can
provide additional pressure support, and delay gas collapse even in
more massive haloes, where thermal pressure is less important.  The
hypothesis that gas has not yet fully collapsed in the haloes will be
supported by our results below on spin misalignment angles and the gas
spin parameters.

\citet{Harford2008} show that the baryon mass fraction increases
inside smaller radii and it reaches a value near the cosmic average at
$R_{\rm vir}$. The $f_{b}$ values in Figure~\ref{b_frac} decrease as
the average volume (and radius) decreases.  In other words, there is
no evidence of an overall concentration baryons near the DM halo
center of mass in our simulations, suggesting that thermal and
turbulent pressure can delay gas collapse even inside 0.5$R_{\rm
  vir}$.

Figure~\ref{b_frac} also shows a large scatter in $f_b$, with
halo-to-halo variations by an order of magnitude.  Despite the low
average $f_b$ at $R_{\rm vir}$, we find three haloes with $f_b$ near
the universal value even at $0.5R_{\rm vir}$. As we will show below,
these haloes are the ones with high central gas spin parameters and
low misalignment angles between gas and DM spin.  

\begin{figure*}
\centering
\includegraphics[height=13cm,width=18cm]{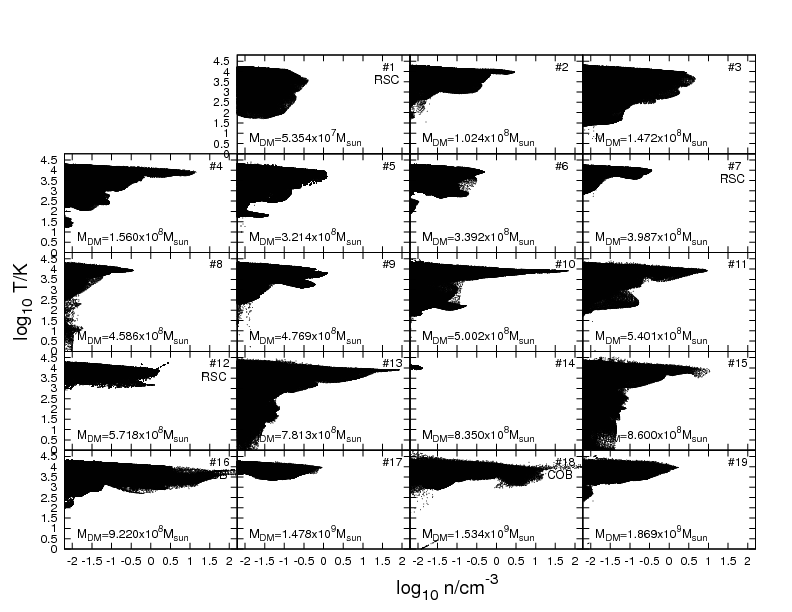}
\caption{Phase diagrams for the gas inside the virial radius in the 19
  simulated haloes.}
\label{planes}
\end{figure*}

In Figure~\ref{planes}, we show the density-temperature $n-T$ phase
diagrams for gas inside $R_{\rm vir}$.  The figure shows that at the
highest densities, most of the gas is heated by shocks and by
compression above the $\sim 10^4$ K limit, but radiative cooling keeps
the gas temperature near this value.  At lower densities, a range of
temperatures is seen, with a significant amount of cooler gas that has
not yet been heated and some gas still significantly above $10^4$K.
Also seen in this figure are a few haloes in which the gas density is
prevented from reaching high values ($n\lsim 1 {\rm cm^{-3}}$).  As we
will see below, some of these haloes have a high degree of rotational
support near their center (including three haloes, \#1, \#7, and \#12,
with full rotational support).

\subsection{Shape of the DM halo and its gas content}

In order to quantify the morphology of both the gas and the DM
components, we computed ratios of the three semi-axis $c\leq b\leq a$
of the ellipsoids, characterized by the eigenvalues of the gas/DM
moment of inertia tensor $I_{ij}$ inside the virial radius $R_{\rm
  vir}$.  The moment of inertia tensor was computed from its standard
definition
\begin{equation}
I_{ij}=\sum_\alpha m_\alpha (r_\alpha\delta_{ij}-r_{\alpha,i}r_{\alpha,j}),
\end{equation}
where $i$ and $j$ label Cartesian coordinates $x$, $y$ and $z$,
$r\equiv\sqrt{r_x^2+r_y^2+r_z^2}$ is the distance from the DM (gas)
center of mass to the $\alpha^{\rm th}$ DM particle (gas cell) and
$\delta_{ij}$ is the Kronecker delta function.

Given the three eigenvalues $\lambda_1\le\lambda_2\le\lambda_3$ of
$I_{ij}$, the semi-axis follow the relations
\begin{equation}
\frac{c}{a}=\sqrt{\frac{\lambda_1}{\lambda_3}},\quad \frac{b}{a}=\sqrt{\frac{\lambda_1}{\lambda_2}}.
\end{equation}

\begin{figure}
\centering
\includegraphics[height=7.9cm,width=7.9cm]{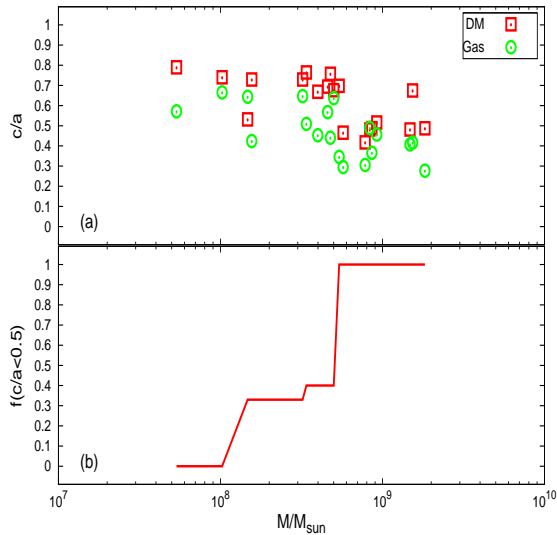}
\caption{(a): The ratio between the semi-minor and semi-major axis
  ($c/a$) for both gas (circles) and DM (squares), as a function of halo
  mass.  The gas component tends to have lower $c/a$ than the DM
  component. (b): The fraction of haloes with gas component $c/a$
  ratio below $0.5$. Despite of the fact that the fraction of haloes
  with $c/a<0.5$ is 100\% above $\sim10^9M_{\odot}$, we find no
  rotationally supported galactic discs above this mass limit.}
\label{c_a}
\end{figure}

Figure \ref{c_a}(a) shows the semi-minor to semi-major axis ratio
$c/a$ as a function of the halo mass. It is clear that the DM
component (squares) tends to be more spherical than the gas component (circles). The
average value is $c/a=0.62$ for the DM and $c/a=0.47$ for the
gas. Figure \ref{c_a}(b) shows the fraction of haloes with the gas
$c/a$ ratio below $0.5$. For masses above $\approx6\times10^8$M$_{\odot}$ all
haloes show $c/a<0.5$. This plot is similar to Figure 3(c) in
\citet{RomanoDiaz2011}, from which they conclude that the disk
fraction is 100\% for more massive haloes.  However, as we will see
below, despite of the low $c/a$, it is far from being
rotationally supported.

\begin{figure}
\centering
\includegraphics[height=7.9cm,width=7.9cm]{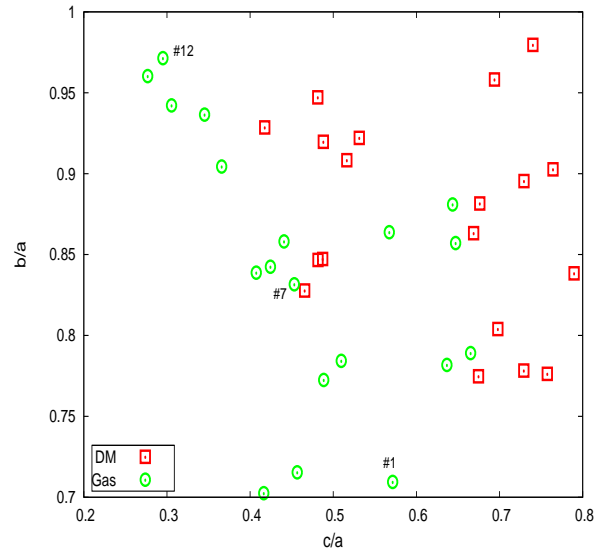}
\caption{The gas (circles) and DM (squares) principal axis ratios in the
  $c/a-b/a$ plane. The DM occupies the high $c/a$, $b/a$ region of
  this plane, while the gas component tends to occupy lower values:
  the DM is more spherical than the flattened gas component. Both
  components typically have a significant triaxiality.}
\label{ellipticity}
\end{figure}

In general, the set of criteria that defines a conventional,
rotationally supported disk should include: (i) low $c/a$ ratio
together with $b/a\approx1$; (ii) alignment of the gas angular
momentum vector with shortest principal axis $c$ and (iii) a gas spin
parameter $\lambda\approx 1$.  Figure \ref{ellipticity} shows the 19
haloes in the $c/a-b/a$ plane.  As this figure illustrates, the DM
component (squares) tends to populate the high $c/a$ ratio region in
the plane, while the gas component (circles) tends to reach much lower
values.  This would be expected if the gas is able to collapse in the
direction perpendicular to the angular momentum vector to assume a
more flattened shape.  We computed the misalignment angle between the
angular momentum and the shortest semi-axis "a" (inside $R_{\rm
vir}$) and found that there is an offset of between 30-90 degrees in
the whole sample.  Overall, the shapes of the DM and the gas are both
typically quite triaxial.  The shapes of the gas and DM nearly remain
closer to prolate rather than oblate ($c/b < b/a$), with average $b/a$
ratios of $b/a=0.84$ for the gas and $b/a=0.87$ for the DM. This
feature has been seen, e.g. in the DM-only simulation of
\citet{Vera-Ciro2011} where the DM matter haloes develop a prolate
shape at the beginning and evolve to an oblate shape at later times.

\subsection{DM halo vs. gas angular momentum}

\citet{Doro70,White1984,HP88} showed that the initial angular momentum
of a system in a hierarchical Universe is proportional to both the
tidal tensor $T_{ij}$ and to the moment of inertia tensor $I_{ij}$
\begin{equation}
J_i\propto\epsilon_{ijk}T_{jl}I_{lk}.
\end{equation}
The evolution of the angular momentum of DM haloes in a cosmological
context has subsequently been revisited by,
e.g. \citet{Vitvitskaetal2002}, to study the importance of mergers.
They found that the angular momentum typically has increases sharply
in major mergers, and declines steadily during periods of gradual
accretion of small satellites.  The angular momentum evolution has
moved from a picture where the tidal forces due to the neighboring
structures are solely responsible for the halo spin to a picture in
which the merger history of the halo has a significant impact.

In order to study the global rotational features of our systems, we
computed the spin parameter $\lambda'$ defined by \citet{Bullock2001}:
\begin{equation}
\lambda_{gas/DM}'=\frac{J_{gas/DM}}{\sqrt{2}M_{gas/DM}\sqrt{G(M_{\rm gas}+M_{\rm DM})R_{\rm vir}}}.
\label{spinparameter}
\end{equation}

The definition of $\lambda'$ in eq. \ref{spinparameter} corresponds to
$1/\sqrt{2}$ times the ratio between the angular velocity $\omega$ and
the angular velocity $\omega_{s}$ of a rotationally supported system
with the same mass and radius (i.e. $M_{gas/DM}$, the gas/DM mass inside
the virial radius).  $J_{gas/DM}\equiv|\vec{J}_{gas/DM}|$ with
\begin{equation}
\vec{J}_{gas/DM}=\sum_{\alpha}m_\alpha \vec{r}_{\alpha,gas/DM}\times \vec{v}_{\alpha,gas/DM}
\end{equation}
is the gas/DM angular momentum. This quantity is computed with respect
to the gas/DM center of mass and the velocities $\vec{v}_{gas/DM}$ are
corrected by the gas/DM center of mass velocity; $m_{\alpha}$ is the
gas/DM mass at position $\vec{r}_{\alpha}$.  All these quantities are
computed inside the virial radius $R_{\rm vir}$ of the
halo.

It has been shown by a number of works (e.g. \citet{Bullock2001} and
references therein), that the spin parameter $\lambda'$ follows a
log-normal distribution:
\begin{equation}
p[\ln(\lambda')]d\ln \lambda' = \frac{1}{\sqrt{2\pi\sigma^2}}e^{-\frac{1}{2}\left[\frac{ \ln(\lambda'/\bar{\lambda'}) }{ \sqrt{2}\sigma }\right]^2} d\ln \lambda',
\end{equation}
where $\bar{\lambda'}$ is defined by
$\ln\bar{\lambda'}=(\sum_i^N\ln\lambda'_i)/N$ (the geometric average
of the spin parameter for $N\gg1$ haloes) and $\sigma$ is its standard
deviation. \citet{Bullock2001} have shown these parameters have values
$\bar{\lambda'}\approx0.04$ and $\sigma\approx0.5$.

In order to study the correlation between the DM and gas spin
orientations, we also computed the misalignment angle $\theta$ between
their spin directions,
\begin{equation}
cos(\theta)=\frac{\vec{J}_{\rm gas}\cdot\vec{J}_{\rm DM}}{J_{\rm gas}J_{\rm DM}}.
\end{equation}

\begin{figure}
\centering
\includegraphics[height=9.5cm,width=\columnwidth]{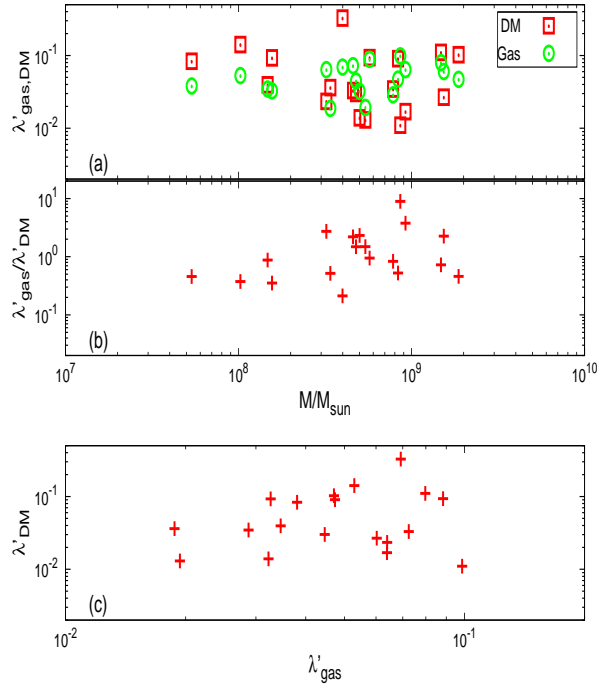}
\caption{(a): The ``global'' spin parameters, measured within $R_{\rm
    vir}$, for both gas (circles) and DM (squares) as a function of
    halo mass. The gas component has lower spin than the DM haloes,
    with a tendency to match at higher masses. (b): the ratio between
    the gas and DM spin parameters as a function of halo mass. The
    ratio between the spin parameters tends to increase at higher
    masses. (c): $\lambda'_{\rm DM}$ vs. $\lambda'_{\rm gas}$. There
    is no clear correlation between these two quantities.}
\label{mass_spin}
\end{figure}

Figure \ref{mass_spin}(a) shows the spin parameters for both gas
(circles) and DM (squares) as a function of halo mass. The average
values over the 19 haloes are nearly identical, $\bar{\lambda}'_{\rm
DM}=0.046$ and $\bar{\lambda}'_{\rm gas}=0.047$.  The average DM spin
value is in good agreement with that found in a previous study using
N-body simulations and focusing on high-redshift haloes
\citep{DavisNatarajan2009}.  Overall, the result $\bar{\lambda}'_{\rm
DM}\sim \bar{\lambda}'_{\rm gas}=0.047$ is also consistent with the
usual simple picture in which the DM and gas initially have similar
spin parameters at the virial scale.  However, we do find an average
ratio of $\langle\lambda'_{\rm gas}/\lambda'_{\rm
DM}\rangle=1.65$. This value is higher than the values reported by
\citet{Gottlober2007} for $\ga10^{14}$M$_{\odot}$ haloes at low
redshift, and also higher than the simple ratio $\langle\lambda'_{\rm
gas}\rangle/\langle\lambda'_{\rm DM}\rangle=1.02$.  This is because
the gas spin parameter tends to be noticeably smaller (by a factor of
$\sim$two) than the DM spin parameter at masses M$_h\la10^8$M$_\odot$.
At higher masses, the gas spin parameters increase and exceed the DM
spin in some cases.  This tendency is illustrated more clearly in
Figure~\ref{mass_spin}(b), where we show the ratio between the gas and
DM spin parameter as a function of the halo
mass. Figure~\ref{mass_spin}(c) shows $\lambda'_{\rm gas}$ v/s
$\lambda'_{\rm DM}$.  There is no a clear trend among these two
quantities. This fact will be supported below by the measurement of
the misalignment angle among gas and DM angular
momentum. \citet{Kimmetal2011} have recently studied the relative
angular momenta of gas and DM haloes, and also found significant
differences. In particular, they found that the discrepancy between
the specific angular momentum of the gas and the DM in the region
$0.1<r/R_{\rm vir}\le1.0$ is due to the cooling process governing the
gas dynamics. The gas is accreted through cold filaments into the
inner regions of the halo, where it condenses; however, the accreted
DM crosses the central region and redistributes the angular momentum
in the outer regions of the halo.

\begin{figure}
\centering
\includegraphics[height=7.9cm,width=7.7cm]{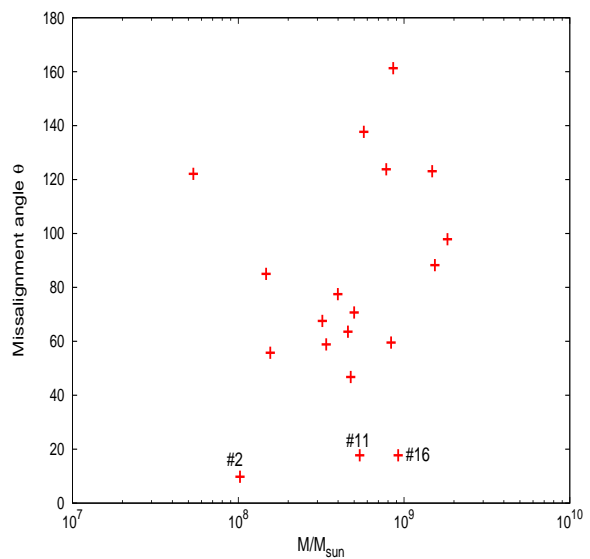}
\caption{Misalignment angle between the DM and gas spin vectors as a
  function of halo mass. There is no clear trend for the misalignment
  with halo mass. This indicates that gas and DM angular momenta have
  different physical origins, likely influenced by the local
  environment rather than determined solely by tidal torques (see
  text for more discussion).}
\label{angle_spin}
\end{figure}

Figure \ref{angle_spin} shows the angle between the DM and the gas
angular momentum vectors. The DM and gas spins are misaligned for all
19 haloes we analyzed, by an average angle
$\langle\theta\rangle=91^\circ$.  This angle is almost 5 times the
average misalignment angle $\theta$ of around $20^{\circ}$ found by
\citet{SharmaSteinmetz2005} at $z=0$, using the SPH technique.  Figure
\ref{angle_spin} shows no trend of $\theta$ with halo mass, and
clearly shows that despite their similar magnitudes, the gas and DM
spins must have different physical origins. Interestingly, we find
that the three most aligned systems have high gas spin parameters and
high baryonic mass fractions. This suggest that the gas in these haloes
has collapsed enough to be aligned.

\begin{figure}
\centering
\includegraphics[height=7.9cm,width=7.7cm]{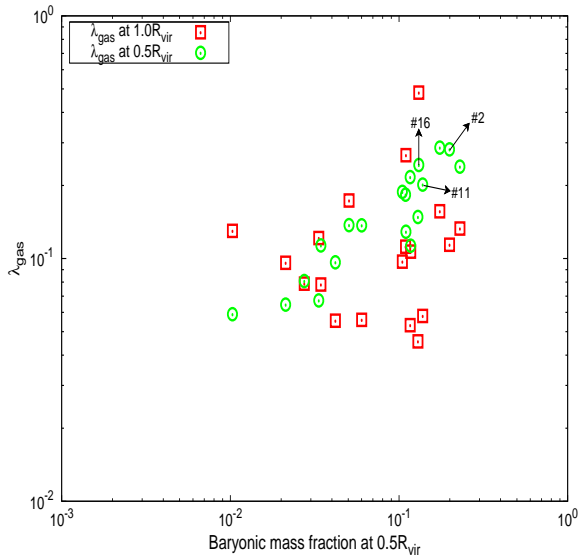}
\caption{Spin parameters $\lambda_{\rm gas}$ at $<R_{\rm vir}$
  (squares) and $<0.5R_{\rm vir}$ (circles) as a function of the
  baryonic mass fraction $f_b$ within $0.5R_{\rm vir}$.  There is a
  very strong correlation between the spin parameter and $f_b$ within
  $0.5 R_{\rm vir}$, showing that the gas inside haloes with more
  collapsed gas eventually reaches higher spins.}
\label{b_frac_spin}
\end{figure}

Figure \ref{b_frac_spin} shows the correlations between the baryonic
mass fraction $f_b(<0.5R_{\rm vir})$, and the gas spin parameter
$\lambda_{\rm gas}$ within both $<1R_{\rm vir}$ (squares) and
$<0.5R_{\rm vir}$ (circles).  There is a clear correlation between
$\lambda_{\rm gas}(<0.5R_{\rm vir})$ and $f_b(<0.5R_{\rm vir})$,
showing that gas inside haloes with more collapsed gas reaches a
higher spin.  In particular, the three haloes with the lowest
misalignment angles present high $f_b$ and $\lambda_{\rm gas}$.
This is consistent with the simple picture in which the DM and gas
have similar angular momenta at larger radii, but the gas can continue
to collapse along their mutual angular momentum axis, and is spun up
by a larger factor - eventually this could lead to disk formation.

\begin{figure*}
\centering
\includegraphics[height=10.5cm,width=18cm]{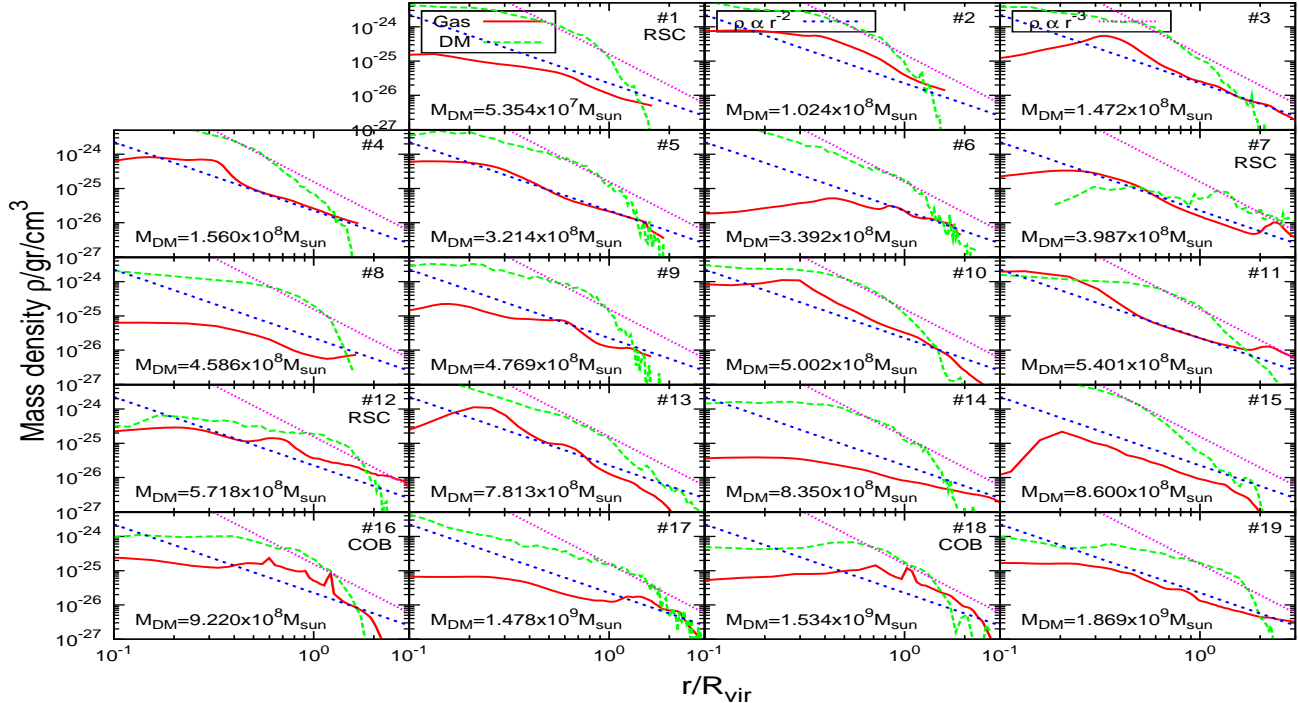}
\caption{Radial mass density profiles for gas (solid line) and DM
  (long-dashed line). For reference, the short-dashed lines and the
  dotted lines show power-law slopes corresponding to an isothermal
  profile ($\rho\propto r^{-2}$) and the NFW profile at large radius
  ($\rho\propto r^{-3}$). The profiles of the most massive haloes
  reveal the presence of over-dense blobs at $\sim R_{\rm vir}$, which
  we argue are candidate sites for the formation of SMBHs. These blobs
  result from large-scale interacting shocks near the halo center. Due
  to the asymmetry of the interaction, the blobs retain a non-zero
  velocity and are able to migrate to the halo outer region (see
  text).}
\label{rho_radial}
\end{figure*}

\begin{figure*}
\centering
\includegraphics[height=9.5cm,width=18cm]{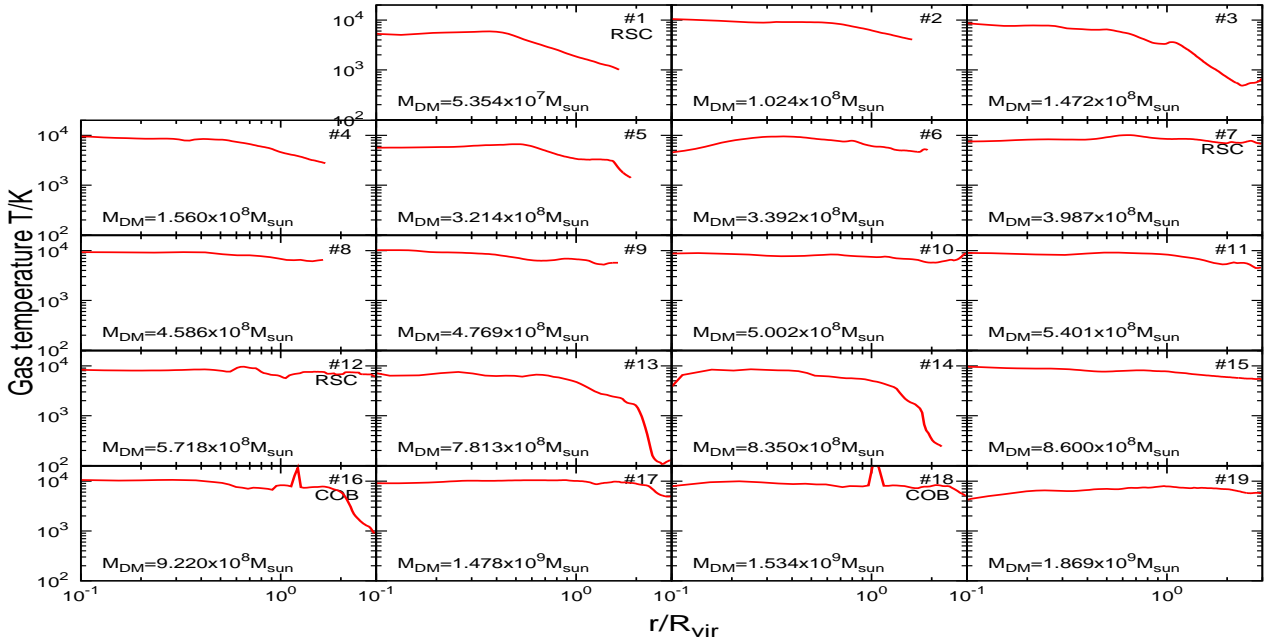}
\caption{Mass-weighted gas temperature profiles. Due to atomic line
  cooling the gas temperature is almost flat at $10^4$K in the central
  regions. The peaks in temperature in two of the most massive haloes
  are associated with the overdensities in Fig.~\ref{rho_radial}, and
  are unphysically large (due to the activation of a numerical
  temperature floor). }
\label{T_radial}
\end{figure*}

\begin{figure*}
\centering
\includegraphics[height=8cm,width=12cm]{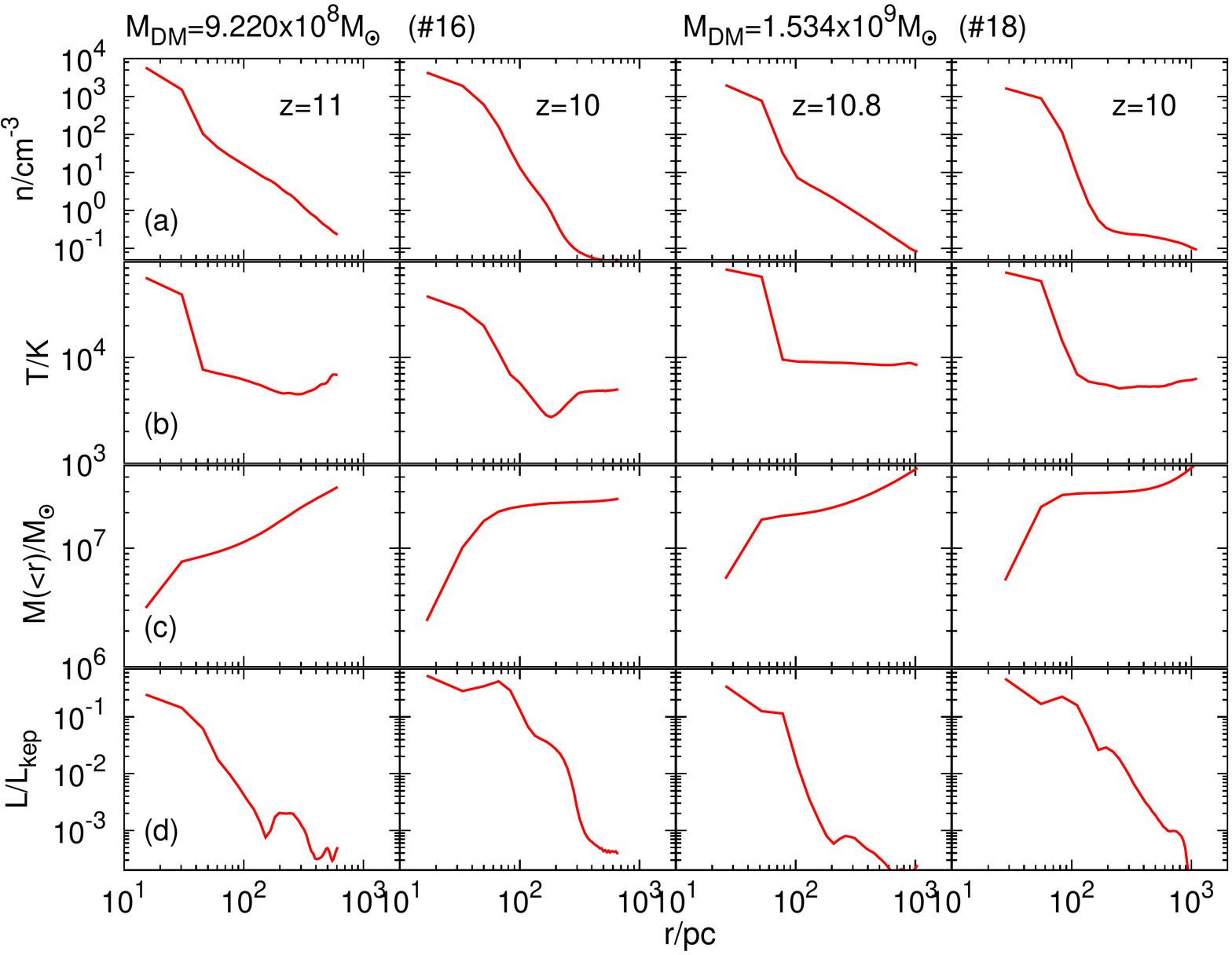}
\caption{Radial profiles at the two overdensities at two different
  redshifts, including the number density (a), the temperature (b),
  the gas enclosed mass (c) and the specific angular momentum normalized
  to the Keplerian angular momentum (d). The first two columns
  correspond to the over density hosted by the
  $M_{\rm DM}=9.220\times10^8M_\odot$ DM halo (\#16) and the last two
  columns correspond to the $M_{\rm DM}=1.534\times10^9M_\odot$ DM halo
  (\#18). The central temperatures are unphysically high because of
  the activation of the numerical temperature floor. Row (d) shows a
  maximum specific angular momentum ratio $L/L_{kep}\approx0.5$. This
  value is similar to the value found by \citet{ReganHaehnelt2009a} at
  $r\approx$ few$\times10$ pc, supporting the notion that these blobs
  are promising sites of SMBHs formation.}
\label{SMBH_radial}
\end{figure*}

\subsection{Radial profiles}
\label{profiles}

We computed the radial profiles of different quantities for both the
gas and the DM component of each halo. This includes the mass-weighted
specific angular momentum and the mass density,
separately for the gas and the DM.  For the gas component, we
additionally computed the mass-weighted temperature.  All quantities
were computed as spherical averages centered at the component (gas/DM)
center of mass.  Note that this, in general, does not coincide with
the point with the highest density.

More specifically, the mass-weighted average value $\bar{q}$ of a
physical quantity $q$ was computed as
\begin{equation}
\bar{q}(\bar{r})=\frac{\sum_{i}q(r_i)m(r_i)}{M_{\bar{r}}},
\label{average}
\end{equation}
where $r_i$ is the position of each DM particle or hydro cell inside
an spherical element of volume given by $\Delta V\equiv 4\pi \bar{r}^2
\Delta r$.  The averaging bin $\Delta r$ takes the value $\Delta r=8
\Delta x$ for gas and $\Delta r=16 \Delta x$ for DM, with $\Delta x$
the minimum resolved scale of the simulation; $m(r_i)$ is the DM or
baryonic mass at the position $r_i$. $M_{\bar{r}}$ is the total DM or
baryonic mass inside the averaging volume and $\bar{r}\equiv r+\Delta
r/2$.

{\em Density profiles.}  Figure \ref{rho_radial} shows both the gas
(solid line) and DM (long-dashed line) density profiles. As mentioned
above, the DM and gas component are centered at their respective
center of mass. This is the reason why the highest densities are not
located at $r=0$. For reference, the short-dashed line shows the
isothermal radial profile $\rho\propto r^{-2}$ and the dotted line
show a $\rho\propto r^{-3}$, the NFW density profile at large radius
\citep{NFW1997}.  These power-law profiles are not good descriptions
of the majority of the profiles (especially for the baryons) but are
good approximations over limited ranges of radii (especially for the
DM profiles in the outer halo near $R_{\rm vir}$).

{\em Temperature profiles.}  The temperature profiles are shown in
Figure \ref{T_radial}.  As expected from ACHs, the gas temperature is
almost constant around $\sim10^4$K inside the virial radius. This
temperature is not affected by the numerical temperature floor in
eq.~(\ref{T_floor}), because at the densities we resolve inside these
haloes, $T_{\rm floor}$ is much lower than $10^4$K. The two exceptions
are the sharp peaks in the temperature profiles of haloes \#16 and
\#18 near $R_{\rm vir}$ (in the bottom row).  These correspond to
strong overdensities (as shown in Fig.~\ref{rho_radial}) which formed
at $z\approx11$. Despite our poor spatial resolution, as we will argue
below, these runaway overdensities are candidate sites for SMBH
formation.  These objects have masses of $M_{\rm
blob}\approx2\times10^7$M$_\odot$ and $M_{\rm
blob}\approx3\times10^7$M$_\odot$ for haloes \#16 and \#18 (with halo
masses of $9.22\times10^8$M$_\odot$ and $1.53\times10^9$M$_\odot$,
respectively).  Interestingly, in both cases, the strong gas
overdensities are located near the virial radius, rather than near the
halo center.  It is tempting to conclude that we are seeing a merger,
and to associate the overdensities with the cores of the merging
partner haloes.  However, this is not the case: the overdensities are
seen only in the gas and not in the DM.  In other words, the dense gas
``blobs'' are not at the center of another DM halo. We have examined
the evolution of the gas distribution over time, and have found that
the overdensities are produced in large--scale, highly supersonic
shocks, and are originally located near the center of the halo. The
gas flows producing the blobs are not perfectly radial toward the
center of the halo, however, and the interacting gas shows a velocity
relative to the halo's center of mass.  The over-dense blob itself,
produced by the strong supersonic shocks, has a significant residual
velocity relative the halo's center of mass. This residual velocity
exceeds the escape velocity from the central region of the halo,
allowing the blob to travel to the outer region of the halo within a
$\sim$dynamical time. A similar phenomenon has been seen in numerical
simulations of star-forming regions \citep{Padoanetal1997}.

To examine these blobs further, in Figure \ref{SMBH_radial} we show
their radial profiles at two different redshifts, re-centered at the
center-of-mass of each blob.  Row (a) shows the gas (number) density
profiles, and row (b) shows the temperature profiles. As mentioned
above, inside $r\sim100$ pc, the gas temperature increases
unphysically, because the activation of the temperature floor (we
emphasize that our results are conservative, in the sense that without
the temperature floor, the blobs should have collapsed further and
reached even higher densities).  Row (c) shows the enclosed gas mass
and row (d) shows the specific angular momentum, divided by the
Keplerian value at each radius.  This ratio is always below unity, but
increases toward smaller radius, reaching a maximum value of
$L/L_{kep}\approx0.5$ at the innermost regions. This value is similar
to the values shown by \citet{ReganHaehnelt2009a} in their Figure~7 at
few$\times10$ pc. As in that paper, these blobs can be similarly argued
to be sites where compact disks, and eventually SMBHs, may form. Due
to the lack of our spatial resolution, however, we can not say more
about the inner structure of these blobs.

We next compute the mass accretion rates (at $z=10$) onto the dense
blobs as follows:
\begin{equation}
\dot{M}_{blob}=4\pi \bar{r}^2 \left( \sum_i \rho_i \vec{v}_i \cdot \hat{r}_i \right),
\end{equation}
where $\vec{r}_i$ is the gas grid position measured from the blob's
center of mass, $r_i$ its modulus, and $\hat{r}_i$ is the unit vector in that direction;
$\bar{r}$ is the radius averaged in a finite shell, $\rho_i$ is the
density at $\vec{r}_i$ and $\vec{v}_i$ is the gas velocity with
respect to the blob's center-of-mass velocity at $\vec{r}_i$.  We have
found $\dot{M}_{blob}\sim0.5$ ${\rm M_{\odot} yr^{-1}}$ at a distance
$r\approx100$ pc. This value should, of course, be taken only as an upper
limit for the rate that would reach any putative SMBH forming near the
center of the COB. This rate is roughly in agreement with the rates
$\sim {\rm M_{\odot} yr^{-1}}$ typically found inside atomic cooling
haloes \citep{OSheaNorman2007,Shangetal2010}.  This latter value is of
order the mass accretion rate $\sim c_s^3/G\propto T^{3/2}/G$ expected
in a self-gravitating gas at temperature $\sim10^4$K.

\begin{figure*}
\centering
\includegraphics[height=12cm,width=18cm]{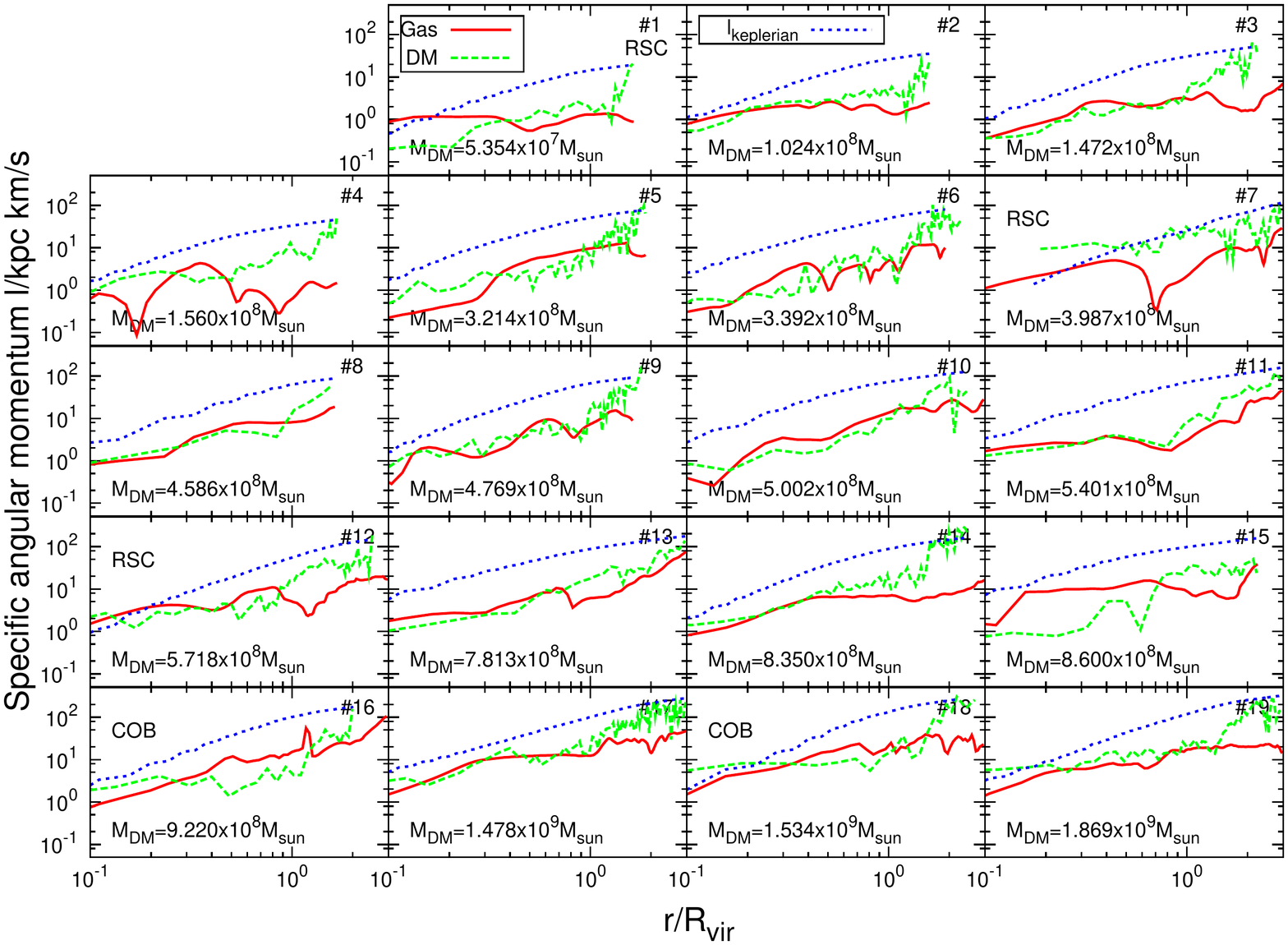}
\caption{Mass-weighted specific angular momentum profiles for the gas
  (solid curves) and DM (long-dashed curves). The gas and DM specific
  angular momenta trace each other and both decrease toward the
  center.  The short-dashed line show the Keplerian specific angular
  momentum (SAM) for the measured density profile. At radii
  $r\ga0.3R_{\rm vir}$ the DM tends to have more specific angular
  momentum than the gas. At smaller radii, there are three cases with
  the gas SAM exceeding the Keplerian values, i.e. the gas in the core
  is rotationally supported.}
\label{L_radial}
\end{figure*}

\begin{figure*}
\centering
\includegraphics[height=12cm,width=18cm]{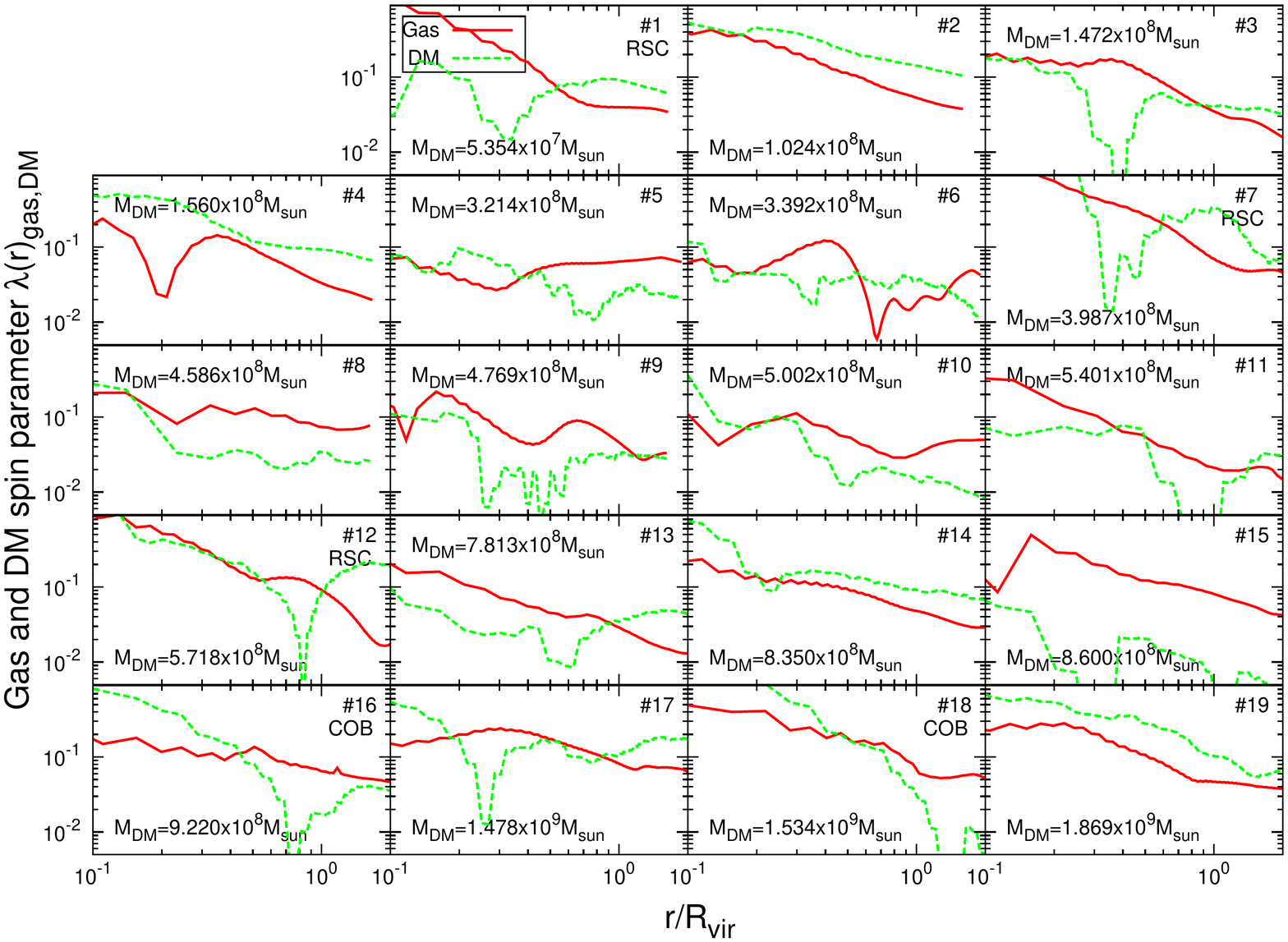}
\caption{Gas and DM spin parameters corresponding to the SAMs in
  Figure~\ref{L_radial}. }
\label{spin_radial}
\end{figure*}

{\em Angular momentum profile.}  Figure \ref{L_radial} shows the
profiles of the specific angular momentum (SAM) for both gas (in solid
line) and DM (in long-dashed line). Overall, the DM and gas SAM trace
each other well at small radii in most of the haloes. This agreement
break down outside $\sim R_{\rm vir}$, where the DM SAM exceeds the
gas SAM, and increases more rapidly with radius, as well.  The
short-dashed line shows the Keplerian SAM profiles, computed for the
density profiles found in the simulations (Fig.~\ref{rho_radial}). At
radii $r\ga0.3R_{\rm vir}$, the Keplerian curves are higher than the
real gas SAM, showing that the gas is not rotationally supported. At
smaller radii there are three haloes (\#1, \#7, and \#12, with $M_{\rm
DM}=5.354\times10^7M_\odot$, $M_{\rm DM}=3.987\times10^8M_\odot$ and
$M_{\rm DM}=5.718\times10^8M_\odot$) in which the gas SAM reaches the
Keplerian value, suggesting the existence of rotationally supported
disks in the inner regions of these three haloes.  However, inspection
by eye does not reveal geometrical disks for these three haloes.  In
the inner 0.1$R_{\rm vir}$, we find axis ratios of $c/a\approx0.6$ and
$b/a\approx0.8$. In particular, the semi-minor to semi-major axis
ratio $c/a$ is close to the ratio $c_s / v_{\rm circ}\approx 0.7$
between the sound speed and the circular velocity; consistent with
thermally supported ``fat disks'' expected when the gas temperature is
not significantly below the virial temperature \citep{OhHaiman2002,ReganHaehnelt2009a}.  
Hereafter we refer to these
objects generically as having rotationally supported cores (RSCs).
Figure \ref{spin_radial} shows the corresponding gas and DM spin
parameter profiles. As shown in Figure~\ref{mass_spin}(b), the gas
spin parameter is higher than the DM spin parameter in some of the
haloes at $R_{\rm vir}$; overall, there seems to be little correlation
between the details of the two.  For most of the haloes, $\lambda'_{\rm
  gas}$ increases at smaller radii.  As seen already in
Figure~\ref{L_radial}, there are three cases in which the gas spin
parameter reaches $\lambda'_{\rm gas}\approx1$, this happens at radii
$r\la0.2R_{\rm vir}$.  The cores of these three haloes within this
radius are rotationally supported.  Interestingly, even in these three
cases, there is a significant misalignment between the direction of
the shortest (semi-minor) axis and the angular momentum vector, by an
angle $20^\circ\la\phi\la40^\circ$ when measured inside $0.3R_{\rm
  vir}$.

\begin{figure*}
\centering
\includegraphics[height=12cm,width=18cm]{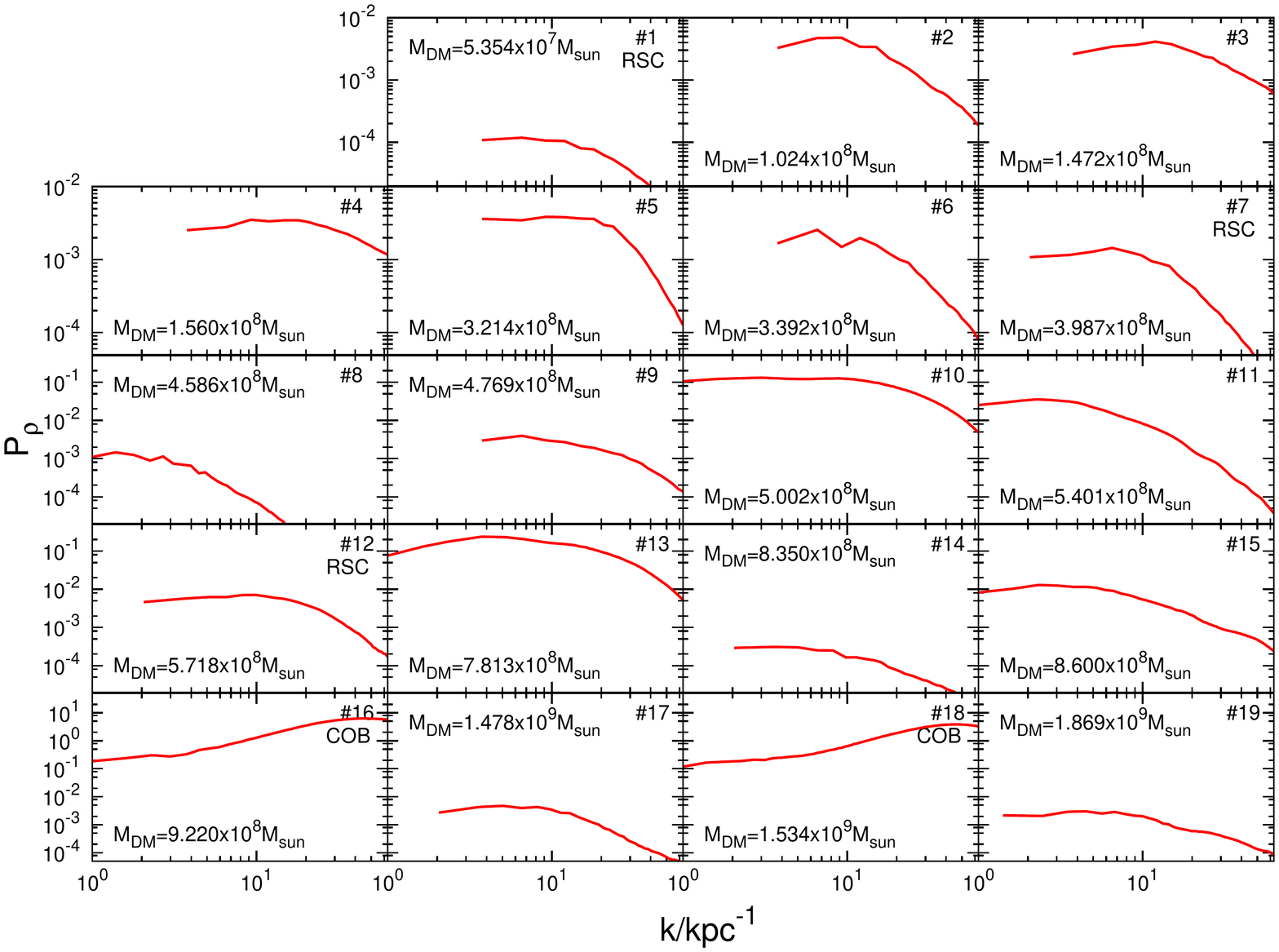}
\caption{Power spectra of the gas density.  For most haloes, the power
  spectra roughly follow power-law scalings of
  $\widetilde{P}_{\rho}\propto k^{\alpha}$ with $\alpha\sim 0$ on
  large scales (expected in the limit of negligible pressure and
  strong shocks) and $\alpha\lsim -2$ (expected in the presence of
  non-negligible pressure and weak shocks) on small scales.  Two of the most massive
  haloes show a clear excess power on small spatial scales (large
  $k$).  These two haloes host over-dense blobs in their outer regions.
  We interpret this to be a result of a highly lumpy and turbulent
  environment, caused by the biased location and the active merger
  history of these two haloes.}
\label{FT_den}
\end{figure*}

\begin{figure*}
\centering
\includegraphics[height=10cm,width=18cm]{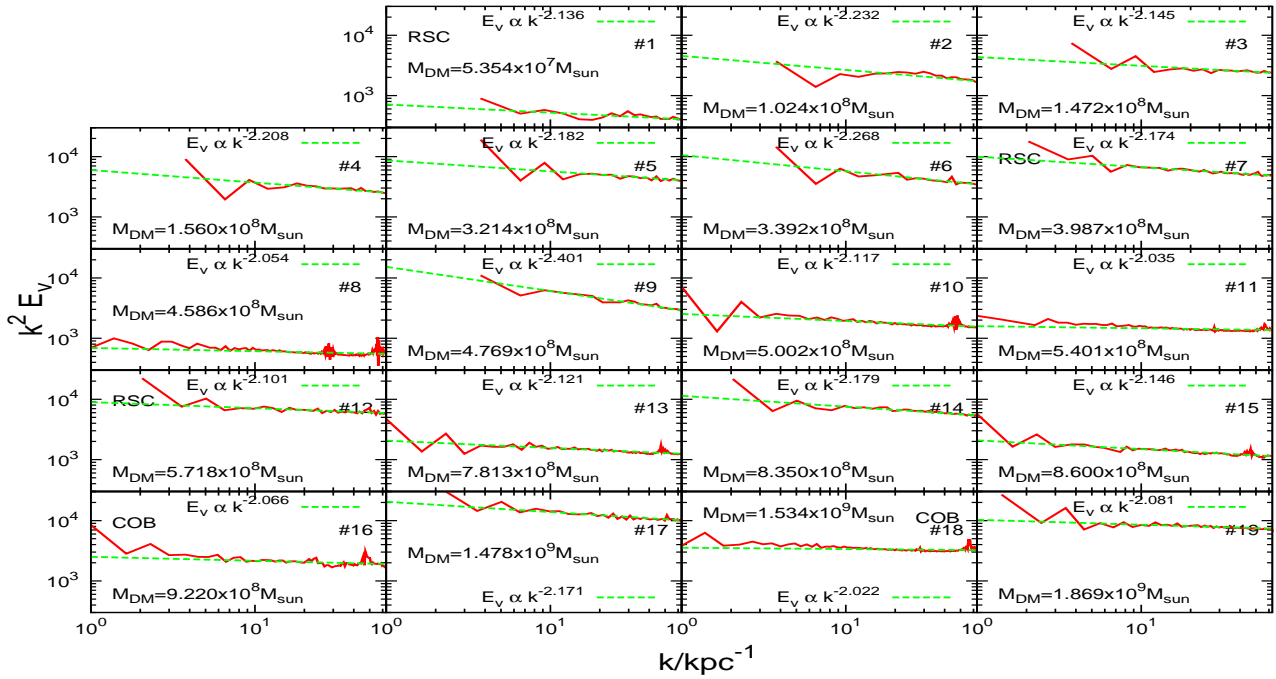}
\caption{Gas velocity power spectra (weighted by $k^2$; solid
  curves). The spectra tend to be very flat.  The long-dashed lines
  show power-law fits $k^2\widetilde{P}_{\rho}\propto k^{\alpha}$ in
  the range 8kpc$^{-1}<k<$80kpc$^{-1}$. The average exponent in this
  range is found to be $\langle\alpha\rangle=-0.15\pm 0.019$
  i.e. closer to a Burgers ($\alpha=0$) spectrum .}
\label{FT_vel}
\end{figure*}

\begin{figure*}
\centering
\includegraphics[height=10cm,width=18cm]{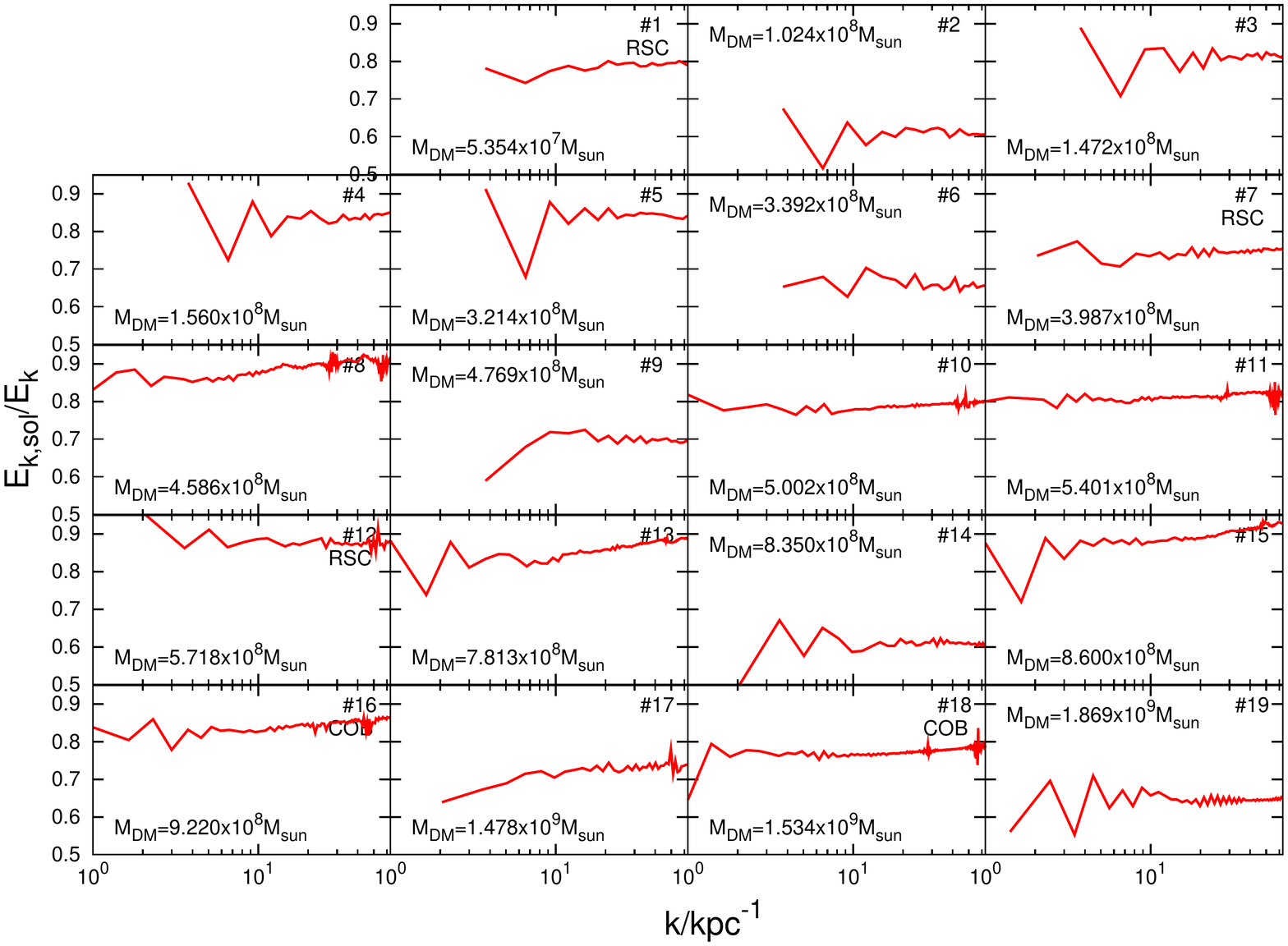}
\caption{The ratio between solenoidal and total (solenoidal plus
  compressional) modes of the kinetic energy. In fully developed
  turbulence, this ratio takes the value $2/3$, because the energy is
  equally distributed among the degrees of freedom (two solenoidal
  modes and one compressional mode).  The ratio $R_k$ is found to be
  $0.6-0.9$ for our 19 haloes.}
\label{FT_ratio}
\end{figure*}

\subsection{Beyond spherically averaged profiles}

\subsubsection{Power spectra of the intra-halo medium}

We next perform a Fourier analysis, in order to understand the
properties of the turbulent environment through which the gas
collapses into the DM haloes. The power spectrum of a physical
quantity is the square of the Fourier transform averaged on spherical
shells of radius $k=\sqrt{k_x^2+k_y^2+k_z^2}$ in wave number space.
Following the above definition we computed the gas mass density $\rho$
power spectrum (PS), in a box of side $s\approx4R_{\rm vir}$ around
the halo's center of mass, as:
\begin{equation}
\widetilde{P}_{\rho}=4\pi k^2 \tilde{\rho}_{k}\cdot\tilde{\rho}_{k}^{*}, 
\end{equation}
and the gas velocity PS
\begin{equation}
\widetilde{E}_{v}=\frac{1}{2}(4\pi k^2 \tilde{v}_{k}\cdot\tilde{v}_{k}^{*}), 
\end{equation}
where $\tilde{v}_{k}$ ($\tilde{\rho}_{k}$) is the Fourier transform of
the gas velocity (density), $\tilde{v}_{k}^*$ ($\tilde{\rho}_{k}^*$)
is its complex conjugate, and $k\equiv|\vec{k}|$ is the modulus of the
wave number. For the velocity field, we have subtracted the motion of
the center of mass, to avoid any spurious turbulence.  The pre-factor
of $1/2$ in the velocity power spectrum converts this quantity into
the kinetic energy stored in wave numbers around $k$.  With this
factor included, we computed the fraction $\widetilde{R}_{k}$ of the
kinetic energy in solenoidal modes,
\begin{equation}
\widetilde{R}_{k}=\frac{\frac{1}{2}(4\pi k^2 \tilde{v}_{s,k}\cdot\tilde{v}_{s,k}^{*})}{\frac{1}{2}(4\pi k^2 \tilde{v}_{k}\cdot\tilde{v}_{k}^{*})} ,
\end{equation}
where the compressional ($\nabla\times \tilde{v}_{c,k}=0$) and
solenoidal ($\nabla\cdot\tilde{v}_{s,k}=0$) components of the
velocity field are given respectively by
\begin{equation}
\tilde{v}_{c,k} = (\tilde{v}_{k}\cdot\vec{k})\vec{k}/k^2
\end{equation}
and
\begin{equation}
\tilde{v}_{s,k} = \tilde{v}_{k}-(\tilde{v}_{k}\cdot\vec{k})\vec{k}/k^2. 
\end{equation}

In turbulent fluid dynamics, there are two well--studied limits for
the turbulent energy power spectrum: the $k^{-2}$ 1D Burgers spectrum
and the $k^{-5/3}$ 3D Kolmogorov spectrum \citep{K41}. The first one
describes a system with velocity $u$ and a fluid viscosity $\nu$
governed by the equation
\begin{equation}
\frac{\partial u}{\partial t}+u\frac{\partial u}{\partial x}=\nu \frac{\partial^2 u}{\partial x^2}.
\label{Burgers}
\end{equation} 
For a high Reynolds number or $\nu\rightarrow 0$ (applicable for this
work), eq.~\ref{Burgers} describes the evolution of a 1D shock
wave. Despite the one-dimensionality of the Burgers equation, it is
often used as a reference for turbulent 3D fluids. A different 3D
treatment comes from the seminal work of \citet{K41}. In this theory, the
statistical properties of a turbulent fluid with viscosity $\nu$, mean
energy dissipation rate per mass $\epsilon$ and a minimum scale $\ell$
associated to the energy dissipation (by viscosity) are fully
determined by these three (or two if $\nu=0$) quantities in the
inertial scale range: $L<l<\ell$, with $L$ the scale associated to the
source of the turbulence (in our case $L\ga R_{\rm vir}$).
  
Figure \ref{FT_den} shows the density power spectra for the gas in the
range of wave numbers corresponding to proper scales $0.1$ kpc$\lsim 2\pi/k \lsim 6.3$ kpc.  
\citet{SaichevWoyczynski1996}
discussed two limits of the scaling relation for the density PS in the
Burgers equation: $\widetilde{P}_{\rho}\propto k^{0}$ and
$\widetilde{P}_{\rho}\propto k^{-2}$. The $\widetilde{P}_{\rho}\propto
k^{0}$ case corresponds to the limit of negligible pressure in which
the fluid is dominated by strong shocks while
$\widetilde{P}_{\rho}\propto k^{-2}$ arises in the presence of fluid
pressure, i.e. a weakly shocked fluid. 
The gas density power spectra in our simulations do not show a
well-defined single power-law, such as found in
\citet{GazolKim2010}. The PS is flatter on large scales, above $\sim
1$ kpc, whereas for most haloes, the power decreases steeply with $k$ on
smaller scales.  The power-law slopes on large (small) scales are
close to $\alpha\approx 0$ ($\alpha=-1.5$).  \citet{KimRyu2005} have
shown that the slope of $\widetilde{P}_{\rho}$ flattens as the Mach
number ${\mathcal M}$ increases.  Following \citet{GazolKim2010}, our
power spectra can be interpreted as evidence for a turbulent cascade
from highly supersonic turbulence (with large ${\mathcal M}$) on the
virial (shock) scales of $\sim$kpc, to low-${\mathcal M}$ turbulence
on small scales (higher $k$).

Probably the most significant feature in these power spectra is that
for two of the most massive haloes, the PS {\em increases} at large
$k$. These behavior is characteristic of systems with self-gravity as
shown by \citet{Collinsetal2012}, where the densest regions have
enough mass to collapse gravitationally and form gas clumps increasing
the power on small scales.  As mentioned by \citet{Collinsetal2012},
this feature in the density PS is indicative of a highly lumpy
environment. In our case, this occurs for the two haloes which
developed over-dense blobs, and clearly distinguishes these two haloes
from the others.  It is worth noting that the excess power is not
simply caused by the presence of these blobs: the wave numbers with
excess power cover a range from $\la100$ pc to $\sim600$ pc, and show no
sign of convergence on small scales.  This means there is excess power
all the way down to our resolution scale of $\approx 10$pc, smaller
than the size $\sim100$pc of the individual over-dense blobs, as well
as on scales much larger than blobs.  Our interpretation is therefore
that the over-dense blobs are a result of a highly lumpy and turbulent
environment.  Below, we will hypothesize that this is a result of
these haloes biased location (at the knots of the cosmic web) and
their merging and cooling history (these haloes are among those with
the largest number of mergers, and could cool their gas early on
$z>16$).

Figure \ref{FT_vel} shows the gas velocity power spectra (compensated
by $k^2$).  This plot is useful in order to differentiate between a
Kolmogorov ($\propto k^{-5/3}$, or $\alpha=1/3$) and a Burgers
($\propto k^{-2}$, or $\alpha=0$) power spectrum.  Interestingly,
Figure~\ref{FT_vel} shows a flat PS for all 19 haloes, with an average
power-law exponent of $\langle\alpha\rangle=-0.15\pm0.019$. In other
words, the gas component for these 19 ACHs is closer to the Burgers PS
(consistent with earlier findings in \citet{jpp2,jpp3}).
Figure~\ref{FT_ratio} shows the solenoidal to total (solenoidal plus
compressional) kinetic energy ratio. In an ideal fully developed
turbulent fluid, this quantity should take the value $R_{k}=2/3$,
showing the equal distribution of energy in two solenoidal,
divergence-free $\nabla\cdot\tilde{v}_{s,k}=0$, modes and a single
compressional, curl-free $\nabla\times\tilde{v}_{c,k}=0$, mode.
Figure \ref{FT_ratio} shows that the distribution of energy typically
exceeds the equipartition value of $2/3$; $R_{k}\approx 0.6-0.9$.
This implies that turbulence has not had time to settle into a steady
pattern.

\begin{figure*}
\centering
\includegraphics[height=13cm,width=18cm]{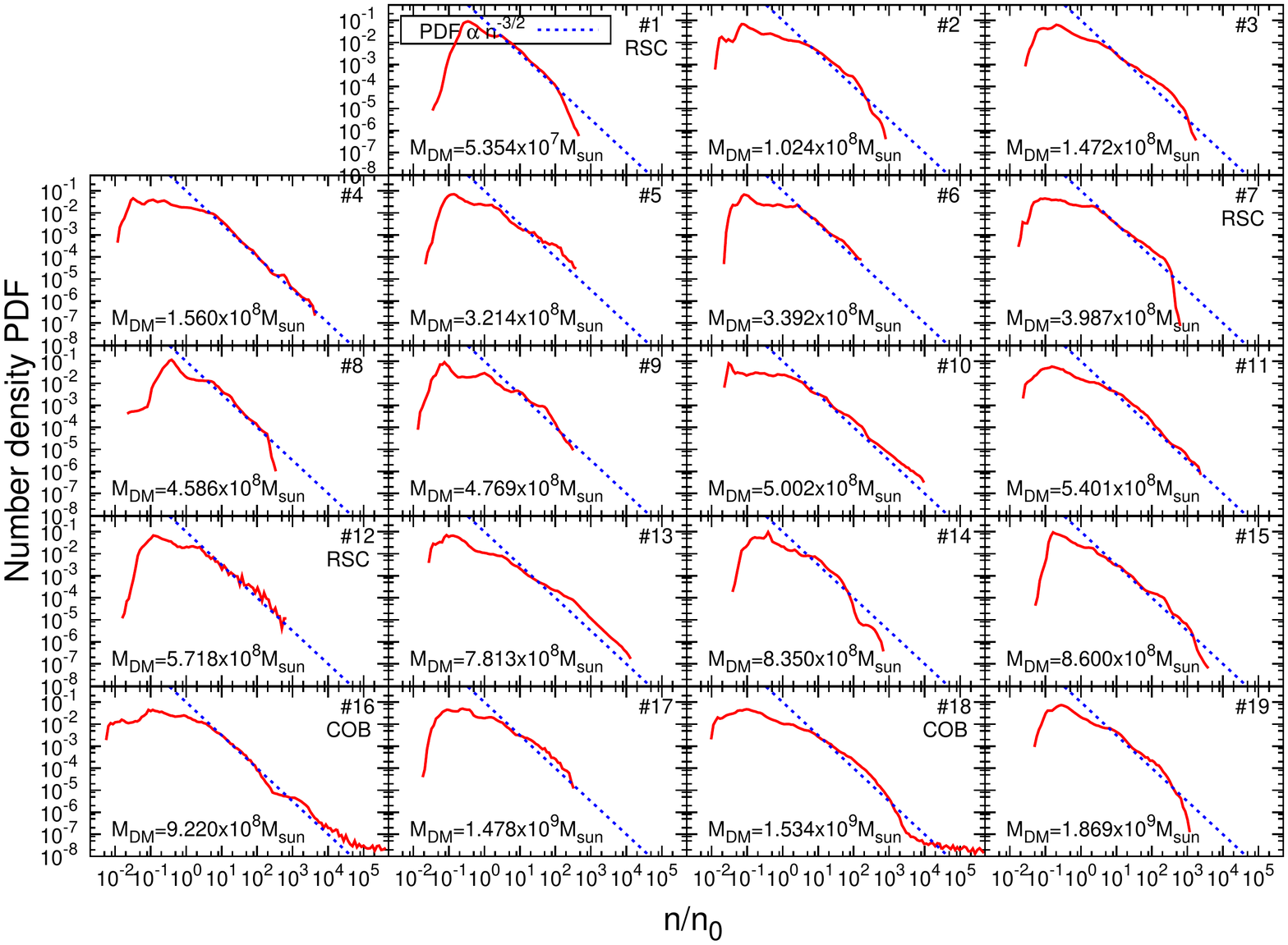}
\caption{Number density PDF (solid curves), normalized by the mean
density $n_0$. The PDFs peak near $n\sim10^{-1}n_{0}$, indicating that
most of the gas remains at low density, as a small fraction of the gas
collapses to high density near the center of the DM halo. Above
$n\sim10n_{0}$, the PDFs develop power--law shapes.  The dashed lines
show a PDF $\propto n^{-3/2}$, characteristic of the first stages of
isothermal gravitational collapse.  Two of the most massive haloes
develop a tail at densities above $\sin10^4n/n_{0}$. These two haloes
host massive over-dense blobs near their outskirt.}
\label{PDF_rho}
\end{figure*}

\begin{figure*}
\centering
\includegraphics[height=13cm,width=18cm]{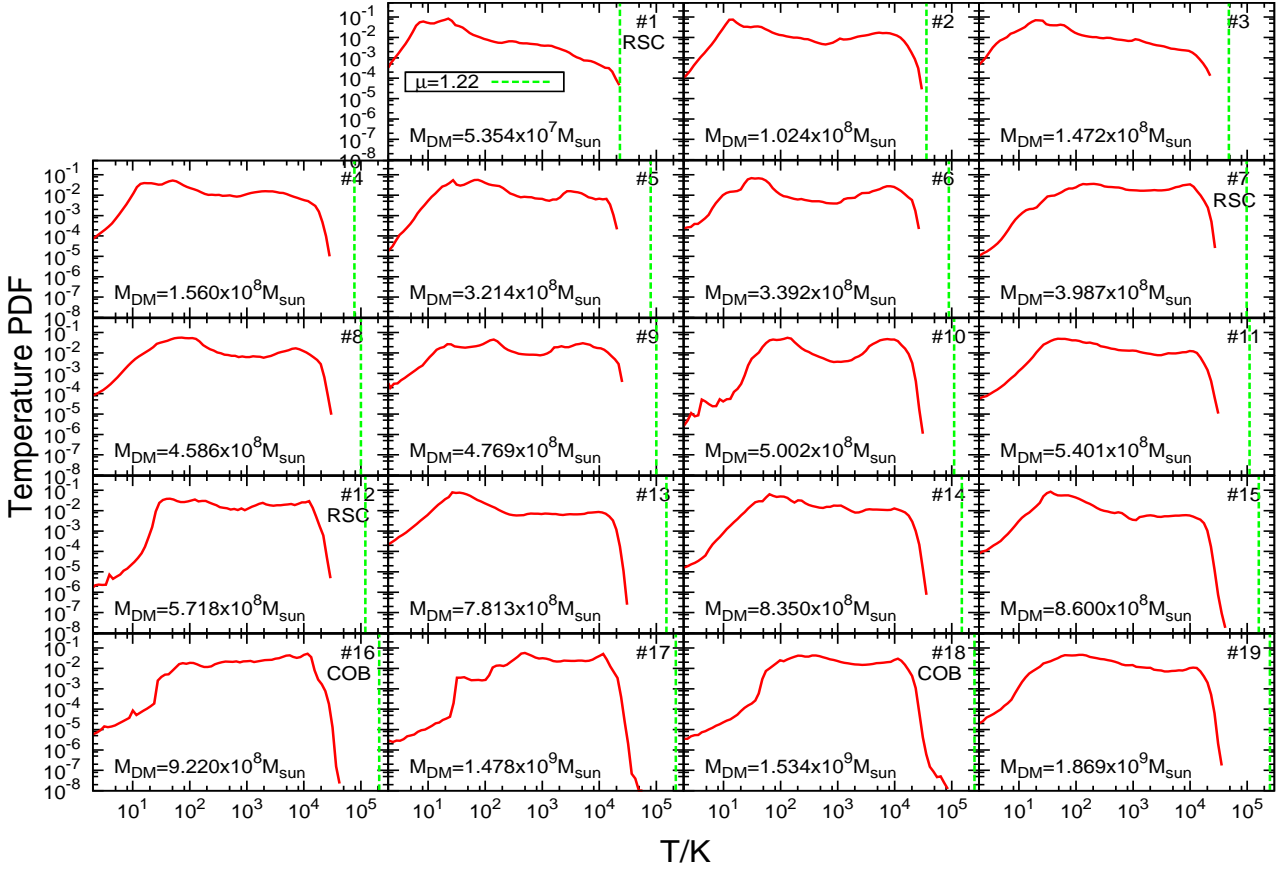}
\caption{Gas temperature PDF (solid curves). The vertical lines show
the virial temperature of each halo. The temperature PDF shows a mixture of hot and cold gas, including (i)
cool, low--density gas far away from the halo center, (ii) hot,
low--density gas, shocked during its collapse, and (iii) hot,
high--density gas in the central region of the halo. The temperature
PDFs are truncated above $\sim10^4K$ by atomic cooling.}
\label{PDF_T}
\end{figure*}

\begin{figure*}
\centering
\includegraphics[height=13cm,width=18cm]{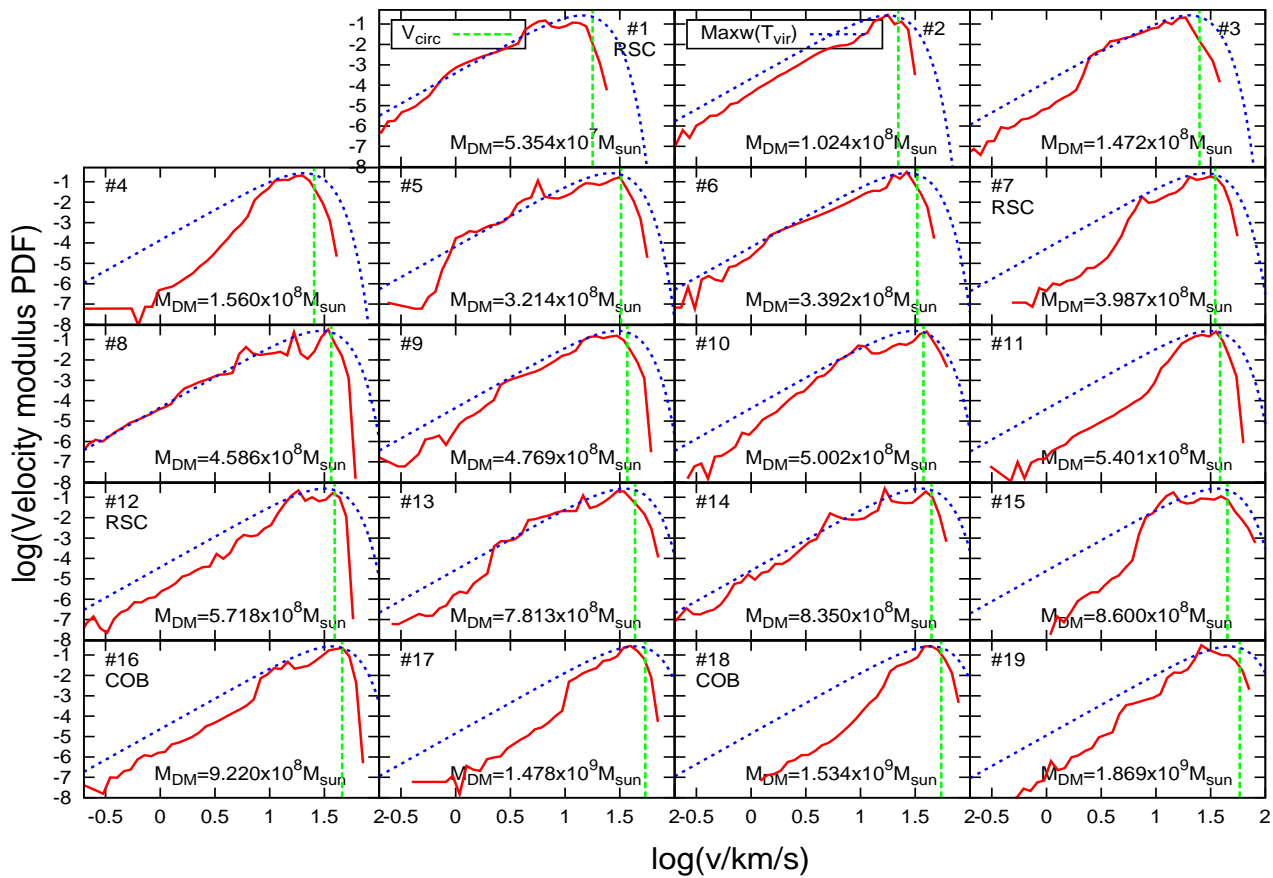}
\caption{Gas velocity PDF (solid line). In most haloes, a peak is
located near the halo's circular velocity (the latter shown by
long-dashed vertical lines). The short-dashed curves show, for reference, Maxwellian velocity
distributions for a temperature $T=T_{\rm vir}$.}
\label{PDF_vel}
\end{figure*}

\begin{figure*}
\centering
\includegraphics[height=18cm,width=2.2\columnwidth]{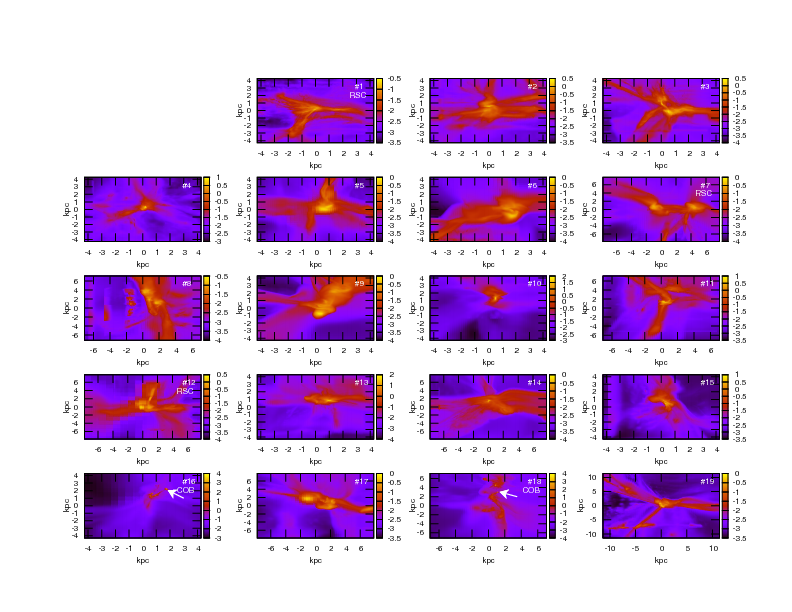}
\caption{The logarithm of the projected mass--weighted
density. Despite significant variations in their details, all haloes
are characterised by a dense core, and filamentary accretion. Two of
the most massive haloes (\#16 and \#18) develop compact over-dense blobs,
here seen to be located near the haloes' outskirts.}
\label{map_den}
\end{figure*}

\begin{figure*}
\centering
\includegraphics[height=18cm,width=2.2\columnwidth]{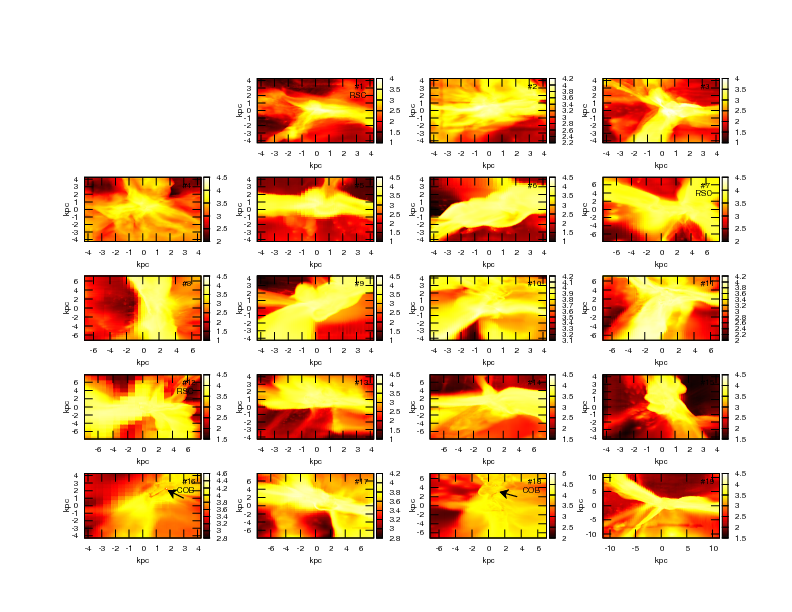}
\caption{The logarithm of the projection of the mass--weighted
temperature. These projected images show filaments converging towards
the central regions of the haloes, producing a shocked and turbulent
environment as shown by the values of the structure functions (see text)}
\label{map_temp}
\end{figure*}

\subsubsection{Probability distribution functions of the IHM}

We next examine the probability distribution functions (PDFs) of the
gas density, temperature, and velocity.  Throughout this discussion,
the PDF is volume-weighted (i.e. $p(q)$ gives the probability to find a
fluid element with a given physical quantity between $q$ and $q +
dq$).

Figure \ref{PDF_rho} shows the PDF of the normalized number density
$n/n_{0}$, where $n_{0}$ is the mean number density inside the volume
analysed (i.e. inside 2$R_{\rm vir}$).  Each of the 19 haloes shows a
distinct peak at $\sim 10^{-1}n_0$. Similar peaks can be seen in
Figure 9 of \citet{jpp3}; in that paper, the peaks show a clear
evolution with time, and become more pronounced at lower redshifts.
As in \citet{jpp3}, the peaks shown in Figure~\ref{PDF_rho} result
from the gas collapsing to the central regions of the DM halo. Through
this process, the gas piles up in a very small percentage of the
volume, while the low density regions fill the rest of the volume.  
At densities above $\sim 10^{1}n_0$, the PDF develops a nearly
power-law shape (the dashed lines in Fig.~\ref{PDF_rho} show a
power-law PDF$\propto n^{-3/2}$ for reference).  This is explained by
\citet{Kritsuk+11} as a feature from the first stages of a nearby
isothermal gravitational collapse.

Two of our 19 haloes developed high-density tails, extending above
$\sim 10^{4}n_0$. These two haloes are the ones that exhibit large
density fluctuations at small scales (Figure~\ref{FT_den}), and
develop the compact over-dense blobs (COBs).  Such high-density tails
result from self gravity acting on the gas, and have been seen in
different contexts in simulations that include self-gravity, such as
isothermal supersonic turbulence (e.g \citealt{Kritsuk+11}),
isothermal MHD turbulence (e.g. \citealt{Collinsetal2012}) and in a
cosmological simulations of primordial gas collapsing in a lower-mass
$\sim3\times10^7$M$_{\odot}$ DM halo \citep{jpp3}. An interesting
feature of these two haloes is that they have the broadest number
density PDFs.  This signature is consistent with these two being the
most shocked and turbulent haloes of the sample.

Figure \ref{PDF_T} shows the gas temperature PDFs for each of the 19
haloes. The vertical dashed lines correspond to the virial
temperatures computed assuming a mean molecular weight $\mu=1.22$
(appropriate for our nearly neutral H+He gas).
The shape of the temperature PDF can be understood as a mixture of
cold and hot gas interacting through the collapse process. All PDFs
show a sharp cut-off above $\sim10^4$K as a consequence of atomic
cooling, and (except for the two smallest haloes \#1 and \#2) the PDFs
do not extend to temperatures as high as $T_{\rm vir}$.  The PDFs all
show cold tails, corresponding to uncompressed gas from the IGM.  The
PDFs for haloes \#16 and \#18 extend close to $\sim 10^5$K; as
mentioned above, this is due to the activation of a numerical
temperature floor in the COBs in these haloes.

Figure \ref{PDF_vel} shows the PDF of the gas velocity modulus, defined as 
\begin{equation}
v=\sqrt{(v_x-v_{x,CM})^2+(v_y-v_{y,CM})^2+(v_z-v_{z,CM})^2}
\end{equation}
where $v_{i,CM}$ is the center of mass velocity of the gas in each
Cartesian direction. For reference, the short-dashed curves show
Maxwell-Boltzmann distributions assuming a gas temperature $T=T_{\rm
vir}$. The velocity PDF resembles features of a Maxwell distribution: for
most of the haloes it has a power-law tail at low velocities that is
well approximated by the slope PDF$\propto v^3$, shows a knee at
$v\approx\sqrt{2k_B T_{\rm vir}/m}$, and has an exponential cutoff at
higher velocities.  For reference, the long-dashed vertical lines
correspond to the halo circular velocity $v_{circ}\equiv(GM_{\rm
  vir}/R_{\rm vir})^{1/2}$.  Interestingly, the peak of the velocity
PDFs are at high velocities, only slightly below (and in some cases
coincident with) $v_{circ}$. This is in contrast with the temperature
PDF, which shows that the gas cools well below the virial temperature.
This indicates that a large fraction of the gas falls into the haloes
supersonically, at speeds near the halo circular velocity.  Typical
Mach numbers of the gas can thus be roughly estimated simply as the
ratio of the halo circular velocity and the gas sound speed ${\mathcal
  M}\equiv v_{circ}/c_s$.  Using the spherical collapse model to
compute $v_{circ}$ for an ACH of mass $M$ and redshift $z$, and
adopting $c_s=c_s(T=10^4$K), we have
\begin{equation}
{\mathcal M}\approx 2\left(\frac{M}{10^8\mathrm{M}_{\odot}}\right)^{1/3}\left(\frac{1+z}{10}\right)^{1/2},
\end{equation}
showing that our haloes should be able to develop a turbulent
environment with Mach numbers ${\mathcal M}\sim 2-4$, and with more
massive haloes having higher Mach numbers.
We computed the Mach numbers inside the virial radius for our haloes,
and for the most interesting cases, i.e. the RSCs and the COBs haloes, 
we found a Mach number ${\mathcal M}\la 4$ and ${\mathcal M}\ga 5$, respectively. 

\subsubsection{Three-dimensional morphologies}

Figures \ref{map_den} and \ref{map_temp} show logarithmic projections
of the mass--weighted gas density and temperature, respectively, for
each halo.\footnote{The fourth column in the second row shows two
haloes, the halo we analyzed (\#7) is the one close to the center of
the image.}  There are significant variations in the details of the
morphologies of our haloes, but all of them are characterized by a
central overdensity, fed through gas filaments, tracing the underlying
DM structure.

The filamentary accretion produces a shocked and turbulent
environment.  As mentioned above, the spherical collapse model
predicts that our haloes have Mach numbers ${\mathcal M}\sim2-4$ and a
direct computation inside the virial radius gives a Mach number range
of $3\la{\mathcal M}\la6$ allowing the development of supersonic
turbulence. One role of the gas filaments in this context is to feed
the chaotic turbulent environment at the central regions of the
haloes.  As shown in Figure \ref{map_temp} (and supported by the
average radial profiles in Figure \ref{T_radial}) due to the atomic
cooling, the overall gas collapse, and the accretion through the
filaments, both proceed supersonically.  As mentioned in previous
sections, two of the most massive haloes develop $M\sim10^7$M$_\odot$
COBs, which can be seen as high density peaks in the bottom row in
Figure~\ref{map_den} ($\#16$ and $\#18$). These two haloes present a
merger history characterized by almost continuous episodes of minor
mergers as can be seen from Table \ref{mergertable}. The multitude of
mergers likely helped to create and maintain a lumpy, turbulent
environment.

\begin{figure*}
\centering
\includegraphics[height=18cm,width=2.2\columnwidth]{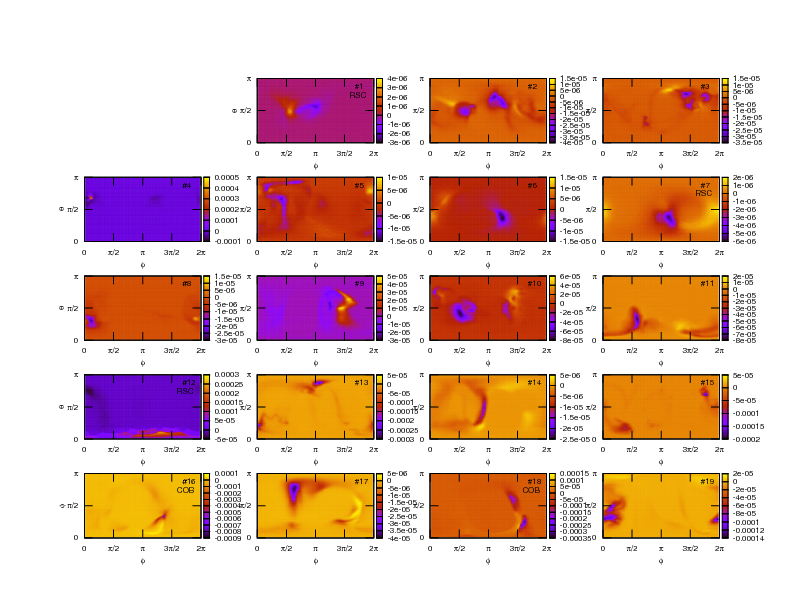}
\caption{Mass flux across the surface of a sphere with radius
$0.5R_{\rm vir}$, centered around each halo.  The mass inflows are
quoted in units of M$_\odot /$yr$/$deg$^2$.  The haloes that develop
compact over-dense blobs (\#16 and \#18) show the highest inward mass
accretion rate (several $\times10^{-4}$M$_\odot /$yr$/$deg$^2$),
$\sim10^1-10^2$ times higher than the accretion rate in haloes with
rotationally supported cores (\#1, \#7 and \#12). In addition, the
mass inflow in the haloes with COBs occurs along the most narrowly
collimated streams. This suggests a causal connection between the
strength of the filamentary accretion and the nature of the object
formed at the center of the halo.}
\label{map_flux}
\end{figure*}

\subsubsection{Turbulence}

The gas in the ACHs have Mach numbers ${\mathcal M}>1$ and they are
thus susceptible to developing shocks and turbulence (see also \citet{Sca} for studies of turbulence in galaxy formation).  Our results
indeed indicate the presence of supersonic turbulence.  There are two
mechanisms that could naturally drive and maintain this turbulence:
(i){\em the filamentary accretion of gas} in relatively narrow,
collimated streams -- the infall speed in these streams is supersonic,
and such streams can collide and create large Mach-number shocks, and
stir the gas near the core of the halo.  (ii) {\em mergers with other
DM haloes} -- mergers can likewise create violent shock waves, and
stir the gas.

To elucidate the role of these processes, we first examine the polar
patterns of mass accretion rate in our haloes. In
Figure~\ref{map_flux}, we show the mass infall rate at each point on
the sphere of radius $0.5R_{\rm vir}$, placed around the
center-of-mass of each halo.  These polar plots show the mass flux (in
units of ${\rm M_\odot~/yr/deg^2}$).  Note that negative values
correspond to inward mass flux (i.e. accretion), and positive values
correspond to outward-directed flows.

This figure shows sharp peaks, indicating that most of the gas inflow
into the haloes occurs along relatively narrow accretion streams.  We
have confirmed that this is true quantitatively, i.e. most of the mass
inflow occurs in a solid angle $\ll 4\pi$. Some haloes (e.g. \#5) have
multiple streams, and some (e.g. \#10) have strong collimated outflows
as well.  Most importantly, this figure shows that the two haloes in
which COBs develop (\#16 and \#18) have the highest mass fluxes by
far. Additionally, the streams feeding gas into these haloes are the
most narrowly collimated (occupying the smallest solid angle).  By
comparison, the haloes having rotationally supported cores present the
lowest mass inflows, and the inflow occurs spread out over a much
larger solid angle (roughly confined, however, to lie within the same
plane).

The difference in mass flux among these two populations is a
factor of $\sim10^{1}-10^{2}$.  These results suggest a causal
connection between the strength of the filamentary accretion and the
nature of the object formed at the center of the halo.  Namely, that
stronger streams lead to stronger shocks and a more turbulent
environment, which can produce COBs. By contrast, the less collimated,
and weaker accretion proceeds in a more orderly fashion, which allows
the specific angular momentum of the gas to be conserved, and a
rotationally supported core (RSC) to develop.

Next, we examine the merger histories of the haloes. In
Table~\ref{mergertable}, for each halo, we list the total number of
mergers it has experienced, the number of minor and major mergers
(defined as having mass ratios below and above $M_1/M_2=1/3$,
respectively), the redshift and mass ratio of each merger, and the
first (highest) redshift at which the gas in the halo could cool.
For our available hydro outputs, $z_{\rm cool}$
was computed as the z when the gas temperature inside $R_{\rm vir}$
reaches the $T=10^4$K. Because we do not have all the hydro outputs
at high $z$ for every simulation, for the $z^*$ cases in the last column 
of Table~\ref{mergertable} we used the DM-only simulations to get this 
redshift. In this case $z_{\rm cool}$ was computed as the $z$ when the halo reaches 
the mass $M=M(T_{\rm vir}=10^4{\rm K})$.

This table shows that the haloes with the COBs are among those with
the highest number of mergers. This is mostly due to their being among
the most massive haloes. However, note that haloes \#17 and \#19 have
comparable masses, but experience fewer (i.e. two and zero) mergers,
and these halo do not develop COBs. This suggests that mergers are
responsible for creating the high mass inflow rates, and causing a
highly turbulent and lumpy environment that leads to strong shocks,
and the formation and spatial displacement of the COBs.  Note that the
COB haloes have not only suffered more mergers, but their gas has also
cooled earlier (see also Fig.~\ref{z_cool} below). This can explain
why in these haloes the gas is more lumpy (as seen in their density
PDFs in Fig.~\ref{PDF_rho}) and can transfer its angular momentum and
fall in rapidly.  Catastrophic transfer of angular momentum and rapid
gas infall of highly clumped gas to small radii is well known to be
associated with early cooling in the low-redshift galaxy formation
literature (e.g. \citealt{Weiletal1998} and references therein).

The same effect can explain the rapid inflow rates we find in our two
COB-forming haloes.
By comparison, in haloes that undergo fewer or no mergers, the
accretion rate remains much lower, spread out over a larger solid
angle, and more coherent.  In this case, a large fraction of the
specific angular momentum can be conserved, leading to rotationally
supported cores.

\begin{table*}
\begin{center}
  \caption{Merging and cooling histories of the 19 haloes in our
    sample. The columns labeled $z_i$ show the redshift of each
    individual merger, with the halo mass ratio given in parenthesis.
    The last column indicates the earliest redshift in which gas
    cooling in each halo was activated.}
\begin{tabular}{ccccccccccccc}
\hline\hline
   &                 &          &           &           &            &           &           &           &           &           &           &\\
Halo &Halo mass        & \#Total  & \#Minor   & \#Major   & z$_{1}$  & z$_{2}$ & z$_{3}$ & z$_{4}$ & z$_{5}$ & z$_{6}$ & z$_{7}$& z$_{\rm cool}$ \\
(number)   &M$_{\odot}$      & mergers  & mergers   & mergers   & (merger    &           &           &           &           &           &           &\\
   &                 &          &           &           &  ratio)& &         &           &           &           &           &\\
   &                 &          &           &           &            &           &           &           &           &           &           & \\
\hline 
   &                 &          &           &           &            &           &           &           &           &           &           & \\
1  &5.35$\times10^7$ & -        & -         & -         & -          & -         & -         & -         & -         & -         & -         & 12.36\\
   &                 &          &           &           &            &           &           &           &           &           &           & \\
2  &1.02$\times10^8$ & -        & -         & -         & -          & -         & -         & -         & -         & -         & -         & 13.94\\
   &                 &          &           &           &            &           &           &           &           &           &           & \\
3  &1.47$\times10^8$ & 1        & 1         & 0         & 10.63      & -         & -         & -         & -         & -         & -         & 13.06\\
   &                 &          &           &           & (0.19)     &           &           &           &           &           &           & \\
4  &1.56$\times10^8$ & 5        & 5         & 0         & 10.60      & 11.24     & 11.85     & 12.39     & 13.26     & -         & -         & 13.10\\
   &                 &          &           &           & (0.08)     & (0.07)    & (0.08)    & (0.14)    & (0.12)    & -         & -         & \\
5  &3.21$\times10^8$ & 5        & 5         & 0         & 10.67      & 11.20     & 12.52     & 13.35     & 14.07     & -         & -         & $>$15.94\\
   &                 &          &           &           & (0.02)     & (0.06)    & (0.07)    & (0.05)    & (0.10)    & -         & -         & \\
6  &3.39$\times10^8$ & 7        & 7         & 0         & 10.45      & 11.03     & 11.57     & 12.35     & 13.10     & 13.77     & 14.86     & 14.89\\
   &                 &          &           &           & (0.02)     & (0.07)    & (0.02)    & (0.05)    & (0.07)    & (0.05)    & (0.08)    & \\
9  &4.78$\times10^8$ & 6        & 5         & 1         & 10.43      & 10.98     & 11.58     & 12.28     & 12.95     & 13.78     & -         & $>$15.98\\
   &                 &          &           &           & (0.02)     & (0.02)    & (0.07)    & (0.04)    & (0.22)    & (0.52)    & -         & \\
10 &5.00$\times10^8$ & 5        & 5         & 0         & 10.68      & 11.25     & 11.74     & 12.55     & 13.37     & -         & -         & 17.58$^*$\\
   &                 &          &           &           & (0.01)     & (0.04)    & (0.09)    & (0.09)    & (0.11)    & -         & -         & \\
12 &5.72$\times10^8$ & 2        & 2         & 0         & 11.55      & 12.28     & -         & -         & -         & -         & -         & $>$15.98\\
   &                 &          &           &           & (0.04)     & (0.05)    & -         & -         & -         & -         & -         & \\
13 &7.81$\times10^8$ & 5        & 5         & 0         & 10.59      & 11.21     & 11.75     & 12.56     & 14.04     & -         & -         & $>$17.42$^*$\\
   &                 &          &           &           & (0.05)     & (0.07)    & (0.02)    & (0.02)    & (0.09)    & -         &           & \\
14 &8.35$\times10^8$ & 1        & 0         & 1         & 12.30      & -         & -         & -         & -         & -         & -         & 14.71$^*$\\
   &                 &          &           &           & (0.75)     & -         & -         & -         & -         & -         & -         & \\
15 &8.60$\times10^8$ & 6        & 6         & 0         & 10.61      & 11.19     & 11.88     & 12.58     & 13.24     & 14.04     & -         & $>$17.12$^*$\\
   &                 &          &           &           & (0.08)     & (0.10)    & (0.04)    & (0.10)    & (0.03)    & (0.05)    & -         & \\
16 &9.22$\times10^8$ & 6        & 6         & 0         & 10.68      & 11.26     & 11.89     & 12.49     & 13.28     & 14.08     & -         & $>$15.97\\
   &                 &          &           &           & (0.03)     & (0.10)    & (0.09)    & (0.07)    & (0.06)    & (0.06)    & -         & \\
17 &1.48$\times10^9$ & 2        & 2         & 0         & 10.50      & 11.04     & -         & -         & -         & -         & -         & $>$15.97\\
   &                 &          &           &           & (0.04)     & (0.11)    & -         & -         & -         & -         & -         & \\
18 &1.53$\times10^9$ & 5        & 5         & 0         & 10.42      & 10.99     & 11.65     & 12.25     & 12.96     & -         & -         & $>$15.87\\
   &                 &          &           &           & (0.10)     & (0.11)    & (0.06)    & (0.08)    & (0.04)    & -         & -         & \\
19 &1.87$\times10^9$ & 0        & 0         & 0         & -          & -         & -         & -         & -         & -         & -         & $>$14.90\\
   &                 &          &           &           & -          & -         & -         & -         & -         & -         & -         & \\
   &                 &          &           &           &            &           &           &           &           &           &           & \\
\hline
\end{tabular}

\label{mergertable}

\end{center}
\end{table*}

\section{Discussion}
\label{Discussion}

\subsection{No disks in atomic cooling haloes}

By definition, the ACHs have a virial temperature $T_{\rm
vir}\ga10^4K$. These haloes lose part of the thermal energy generated
through the gravitational collapse process by H (and He near
$\sim10^5K$) emission line cooling, and reach a temperature floor of
$T\approx10^4K$. This fact can be seen in the temperature profiles in
Figure~\ref{T_radial}, which show a constant $T\approx10^4K$ inside
$\sim R_{\rm vir}$. The gas density profiles approximately follow a
power--law $\rho\propto r^{-2}$, in-between the profiles expected from
pressureless cosmological infall $\rho\propto r^{-1.5}$ and infall of
non-radiating gas through shocks $\rho\propto r^{-2.25}$
\citep{Gunn,Bert}.

One of the main motivation for this work was indeed to look for large
($\sim 0.1R_{\rm vir}$), rotationally supported gaseous disks inside
the ACHs. This is expected in the simplest toy models
(e.g.\citealt{Moetal1998,OhHaiman2002}), in which efficient cooling
allows the gas to collapse preferentially in the direction parallel to
its angular momentum vector, while acquiring increasing rotational
support in the plane perpendicular to this direction -- eventually
settling to a rotationally supported disk.  Interestingly, we find
that collapse indeed occurs preferentially along specific directions,
and the gas has a less spherical shape than the host DM haloes. haloes
with masses above $\sim6\times10^8M_{\odot}$, $100\%$ present a
flattened gas distribution, with $c/a<0.5$.  This is in agreement with
the earlier results by \citet{RomanoDiaz2011}.  However, we find that
the minor axis of the gas distribution is {\em not} aligned with its
angular momentum vector - the offset is between 30-90 degrees. This
invalidates the above simple picture.

Furthermore, we find that none of these flattened gas distributions
are rotationally supported; their spin parameters are well below
unity.  Interestingly, in 3 of our haloes the gas in the core does reach
$\lambda'_{\rm gas}\approx 1$, but these rotationally supported cores
are found in the lowest-mass haloes, and are closer to spherical blobs
than to geometrical disks.  In summary, we have proposed the following
requirements for the gas distribution to qualify as a conventional
``disk'':
\begin{itemize}
\item a low $c/a\lsim 0.5$ and a high $b/a\approx1$, 
\item a high gas spin parameter $\lambda'_{\rm gas}\approx1$ and 
\item an alignment (to within $\sim 20^\circ$) of the angular momentum
vector of the gas with the minor axis of the associated ellipsoid.
\end{itemize}
We find {\em no} disks in our sample of 19 atomic cooling haloes that
satisfy this set of intuitive criteria.

\subsection{Lack of correlation between gas and DM spin}

Another interesting result from our study is that
the directions of the angular momentum vectors of the gas and DM
components are significantly misaligned. The misalignment angle
$\theta>20^\circ$ for most of the haloes; in fact, for a third of the
haloes in our sample, we find $\theta>90^\circ$, i.e. the gas and the
DM components counter-rotate.

Following the simple scenario in which the baryonic matter acquires an
average SAM $J_{\rm gas}/M_{\rm gas}\propto J_{\rm DM}/M_{\rm DM}$, as
assumed in \citet{Moetal1998} and \citet{OhHaiman2002}, we should
expect a negligible misalignment angle.  Three of our haloes present
alignment to within $\theta\approx20^\circ$. As mentioned above, these
haloes have a high spin parameter and a high $f_b$, suggesting that
they are in a more advanced state of collapse compared with the rest
of the haloes.  One reason for the high misalignment angles in most of
the haloes could be that the gas in these haloes have not yet fully
collapsed, and the DM and gas components have not had enough time to
interact and reach a stable, aligned state. As mentioned above,
\citet{Kimmetal2011} have recently studied the relative angular
momenta of gas and DM in haloes at lower redshifts, and also found
significant differences, due to the very different dynamics of the gas
and the DM once they cross the central region of the halo. It has also
been pointed out by Angles-Alcazar \& Dav\'e (in preparation).

\begin{figure}
\centering
\includegraphics[width=1.05\columnwidth]{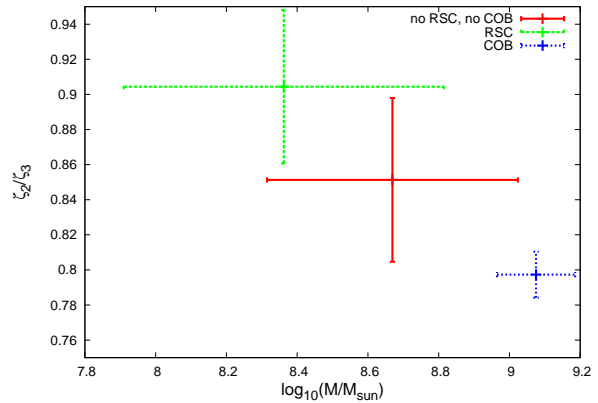}
\caption{The ratio of the power-law exponents of the longitudinal
structure functions of $2^\circ$ and $3^\circ$ order
$R=\zeta_2/\zeta_3$. This ratio is expected to be $\approx 0.74$ for
fully developed, highly supersonic turbulence, and higher for
less-developed, weaker turbulence (see text). Halo type (i.e. haloes
with COBs or RSCs) correlate with both this value and mass of the DM
halo, hinting at turbulence as a possible explanation of the different
morphologies of these objects. The error bars correspond to the r.m.s
among the different haloes for each class of objects.}
\label{fig:structure}
\end{figure}

\begin{figure}
\centering
\includegraphics[width=1.05\columnwidth]{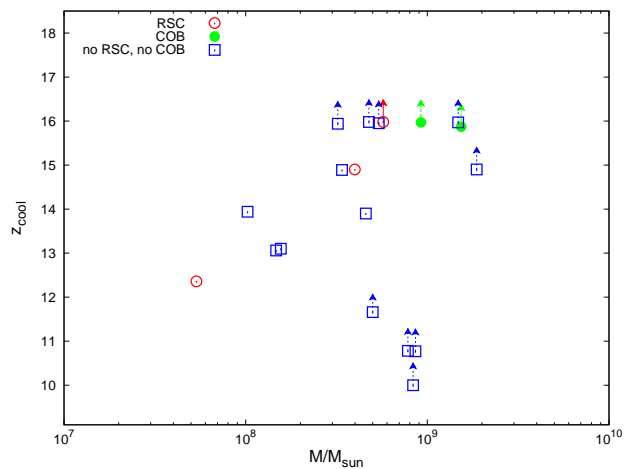}
\caption{The earliest redshift at which atomic cooling is activated in
each of the $19$ haloes. The haloes forming COBs are among the most
massive at $z=10$ and are the first to cool.}
\label{z_cool}
\end{figure}

\subsection{Rotationally supported cores vs. compact over-dense blobs}

Our simulations show evidence for the formation of RSCs and COBs. The
haloes in which these two different gas configurations are found have
very different physical features, in terms of turbulence, merger
history and larger-scale environment.  These difference can help us
understand the reason for having such different primordial objects.

The radial profiles of the specific angular momenta and the spin
parameters show that in the 3 haloes where the gas component reaches a
rotational support at small radius, $\lambda'\approx1$ and a
$L/L_{kep}\approx1$ inside $r\approx0.1R_{\rm vir}$. On the other hand, the
haloes forming COBs have low gas spin parameters, and much flatter
spin parameter profiles, indicating that the gas could efficiently
transport its angular momentum.  As described above, the two types of
haloes have very different merger histories: Table \ref{mergertable}
shows that the COB haloes have suffered many more mergers than the RSC
haloes and that they have cooled their gas earlier
(Fig.~\ref{z_cool}).  The early cooling and the multitude of mergers
can explain the fact that the gas in the COB haloes have a much more
clumpy distribution (Fig.~\ref{PDF_rho}), which may explain the
efficient transfer of angular momentum, and the rapid and narrowly
collimated gas inflows (Fig.~\ref{map_flux}) in these haloes -
creating strong shocks and driving strong turbulence in the central
regions, eventually forming the COBs.

\begin{figure*}
\centering
\includegraphics[height=10cm,width=2\columnwidth]{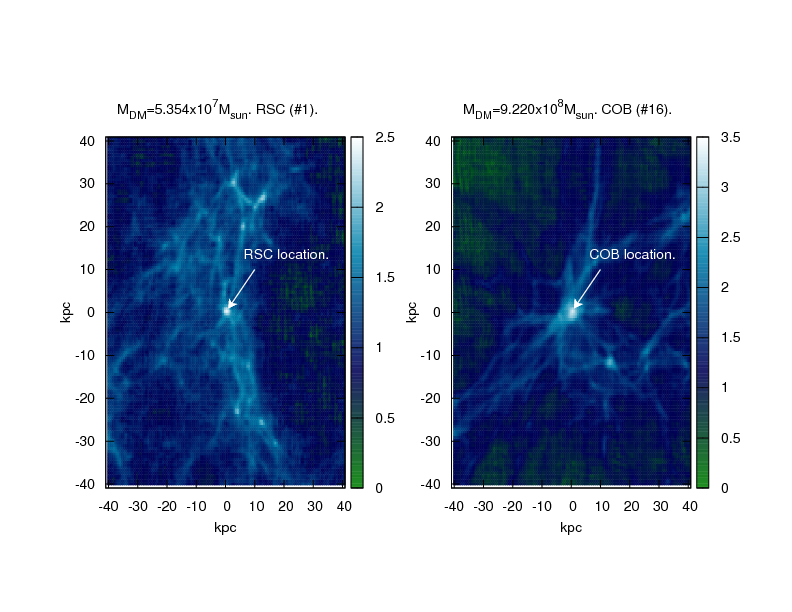}
\caption{Maps of the projected DM surface density, in arbitrary units,
to illustrate the large-scales cosmic environments of one of the haloes
developing a rotationally supported core (RSC, left panel) and a
compact over-dense blob (COB, right panel). The RSC halo is located at
a typical point in a filamentary region, and the COB-forming halo
resides at the center of a dense knot in the cosmic web. The merging
histories of these two haloes are accordingly also very different (see
Table \ref{mergertable}).}
\label{map_DM}
\end{figure*}

In order to distinguish haloes with a diagnostic that is explicitly
related to turbulence, we have measured the velocity structure
functions of the gas within the inner $0.5 R_{\rm vir}$.  The
structure functions of order $p$ are defined as:
\begin{equation}
S_p(l) = \langle |u(x + l) - u(x)|^p \rangle \propto l^{\zeta (p)}
\end{equation}
where the velocity component $u$ is parallel (longitudinal structure
function) or perpendicular (transversal structure function) to the
vector $\vec{l}$, $l\equiv\mid\vec{l}\mid$, and the spatial average is
over all values of the position $x$ (for more details, see
\citet{jimenezboldyrev}). The ratio of the power-law exponents of the
longitudinal structure function of $2^\circ$ and $3^\circ$ order,
$R=\zeta_2/\zeta_3$, has been shown to depend on the strength of
turbulence \citep{jimenezboldyrev}. In particular, for highly
developed supersonic turbulence, \citet{jimenezboldyrev} finds the
general scaling 
\begin{equation}
\frac{\zeta(p)}{\zeta(2)} = \frac{p}{9} +1 - (1/3)^{(p/3)},
\end{equation}
which for $p=2$ yields a value of 0.74.  We have computed this ratio
for each of our 19 halos, and show the average values, with r.m.s.
errors, for the three types of halos separately in
Figure~\ref{fig:structure}.  Interestingly, the two types of halos
with RSCs or COBs have significantly different ratios, $R_{\rm
RSC}=0.90\pm0.04$ versus $R_{\rm COB}=0.79\pm0.01$, which can be
attributed to the different turbulent conditions.  The value for the
COB-forming halos is close to the Boldyrev-type (i.e. highly
supersonic, fully developed) turbulence. From Figure~2 in
\citet{jimenezboldyrev}, we see that the higher value of $R=0.9$ for
the halos with RSCs corresponds to less supersonic, and less
well-developed turbulence (but still above the value for Kolmogorov
subsonic turbulence).  Perhaps most interesting is the clear
correlation between the halo mass and structure function ratio, as can
be seen in Fig.~\ref{fig:structure}.  Without further simulations and
analysis, it is unclear what causes this correlation, but we
speculate, consistent with our interpretations above, that it is due
to the different merger histories, which in turn set the different
turbulent conditions, which in turn create different types of
objects. We will explore this hypothesis in more detail in a
forthcoming publication.

As mentioned above, the mergers suffered by the progenitor halo is not
the only possible source of turbulence. A second possible source is
rapid gas accretion through filaments, following the underlying DM
structures. These two mechanisms can, in principle, both
simultaneously contribute to driving turbulence.  In particular, we
found that the gas accretion rate for the COB haloes is
$\sim10-10^2$ times higher than the one presented in the RSC
haloes, suggesting that this second mechanism must play an important role. 

Finally, it is interesting to ask whether the larger-scale
environments of the haloes may affect their turbulent properties, and
the type of object that forms the haloes. Figure \ref{map_DM} shows
maps of the projected DM surface density 
around the location of one of the RSC (left panel) and COB (right
panel) haloes. The differences are evident: while the RSC halo is
located at a typical point in a filamentary region, the COB-forming
halo resides at the center of a dense knot in the cosmic web. These DM
density maps support the idea that the more numerous mergers and more
rapid accretion flows found in the COB haloes are connected to its
larger-scale cosmic environment.

\section{Summary and Conclusions}
\label{Conclusions}

In this study, we have presented the largest statistical cosmological
hydro-simulation study of atomic-cooling haloes (ACHs).  We studied
the physical properties of 19 isolated haloes at $z=10$, covering a
total comoving volume of $\sim2000$ Mpc$^3$.  The simulations reach a
spatial resolution of $\Delta x\sim10$ (proper) pc.

We presented a comprehensive analysis of the gas and DM physical
properties, including baryonic mass fractions, gas and DM shapes, gas
and DM spin parameters, as well as the radial profiles, power spectra,
and probability distribution functions of various halo and gas
properties. Our main goal was to study and understand the formation of
any large galactic disks among ACHs, which may be stable and avoid
fragmentation, suitable for forming SMBHs. In particular, our aim was
to assess whether disk formation was common among these haloes.

Our main conclusions can be summarised as follows:

\begin{enumerate}

\item None of our 19 haloes has formed a rotationally supported
  geometrical disk.  While haloes with masses above $\sim10^9M_\odot$
  have flattened geometrical shapes, they do not have significant
  rotational support.  Conversely, we find three lower-mass haloes in
  which a rotationally supported core develops, but the shape of this
  core is closer to spherical than to a disk.  Nevertheless, these
  cores are candidate sites for SMBH formation.

\item The gas spin parameter $\lambda'_{\rm gas}$ is higher than the
  DM spin parameter $\lambda'_{\rm DM}$ for some of our haloes above
  $\sim10^8$M$_\odot$.  This result is in superficial agreement with
  the simple picture of galactic disk formation, in which specific
  angular momentum is conserved, and the gas spin parameter increases,
  as the gas collapses inside the DM halo.  However, we find that the
  magnitude of the gas and DM spins have little correlation, and their
  directions are also significantly misaligned. This indicates that
  the gas angular momentum does not follow the above simple picture.

\item Two of our most massive haloes developed very massive compact
  over-dense blobs, with masses of $\sim$few$\times10^7M_\odot$. The
  COBs form near the center of the haloes by strong shocks at the
  intersection of converging filamentary flows. However, unexpectedly,
  the blobs have large residual velocities, and they migrate to the
  haloes' outskirts.  The blobs are self-gravitating, and accrete
  rapidly ($\sim0.5$M$_{\odot}$yr$^{-1}$, measured at a distance of
  $r\approx100$ pc from their center). These COBs show no sign of
  fragmentation at our resolution, and are alternative candidate
  sites for SMBH formation by direct collapse.

\item The intra-halo medium is turbulent and supersonic (with Mach
  numbers ${\mathcal M} = 2-4$). To demonstrate this we have computed
  structure functions and showed that for all haloes they are all
  consistent with structure functions above the Kolmogorov regime.

\item The COB formation is associated with a highly turbulent,
  supersonic (Mach numbers ${\mathcal M}\ga 5$) and lumpy
  environment due to a high mass accretion rate and an active merger
  history. This is likely related to their location at dense knots of
  the cosmic web. On the other hand, the rotationally supported cores
  are associated with haloes that are less turbulent, undergo mergers
  much less frequently, and have a lower gas accretion rate. They are
  located midway along filaments in the cosmic web.

\end{enumerate}

We conclude that the angular momentum transport and the fate of gas in
$z=10$ ACHs is intimately related to the turbulent conditions, which
in turn is determined by the merging history and larger-scale
environment of the halo. We hypothesize that haloes in highly biased
regions have more numerous and earlier mergers, more lumpy and
turbulent gas distributions, and more efficient transfer of angular
momentum. The gas in such haloes can be accreted rapidly, along narrow
streams, into a very dense central object, whose formation involves
strong shocks. On the other hand, haloes in less biased regions have
fewer mergers, and a less turbulent intra-halo medium, where gas
inflow is more coherent and weaker overall. In such haloes, a larger
fraction of the angular momentum can be preserved, and thus eventually
a rotationally supported configuration can be reached.

Our results are based on simplified simulations. Since we assume the gas has primordial
composition and ${\rm H_2}$-cooling and prior star-formation in the
haloes have been suppressed, our simulations can be regarded as
numerical experiments. They may nevertheless be an accurate
description of a small subset of atomic cooling haloes, exposed to
strong enough Lyman-Werner radiation to prevent prior star-formation.
If verified in a larger sample of haloes and with additional gas
physics to account for metals, star-formation and feedback, our
results will have implications for observations of the
highest-redshift galaxies and quasars with {\rm JWST}.

\section*{Acknowledgments}

We thank Greg Bryan, Ryan Joung and Matt Turk for useful discussions.
ZH acknowledges support from NASA grant NNX11AE05G.

\end{document}